\documentclass{emulateapj}
\usepackage{epsfig}
\usepackage{lscape}

\newcommand{\etal}{\mbox{et al.}}
\newcommand{\ergcms}{\mbox{erg cm$^{-2}$ s$^{-1}$}}

\newcommand{\ergsec}{\mbox{erg s$^{-1}$}}

\newcommand{\phcms}{\mbox{photons cm$^{-2}$ s$^{-1}$}}
\newcommand{\degree}{$^\circ$}
\newcommand{\msun}{$M_{\odot}$}

\newcommand{\chandra}{{\it Chandra}}

\newcommand{\sgrastar}{\mbox{Sgr A$^*$}}

\newcommand{\numsrc}{34}

\shortauthors{Muno \etal}
\shorttitle{Diffuse X-ray Emission}

\begin{document}
\title{A Catalog of Diffuse X-ray-Emitting Features within 20 pc of Sgr A$^*$: Twenty Pulsar Wind Nebulae?}
\author{M. P. Muno,\altaffilmark{1,2} F. K. Baganoff,\altaffilmark{3}
W. N. Brandt,\altaffilmark{4} M. R. Morris,\altaffilmark{1} and 
J.-L. Starck\altaffilmark{5}}

\altaffiltext{1}{Department of Physics and Astronomy, University of California,
Los Angeles, CA 90095; mmuno@astro.ucla.edu}
\altaffiltext{2}{Space Radiation Laboratory, California Institute of 
Technology, Pasadena, CA 91104}
\altaffiltext{3}{Center for Space Research,
Massachusetts Institute of Technology, Cambridge, MA 02139}
\altaffiltext{4}{Department of Astronomy and Astrophysics, 
The Pennsylvania State University, University Park, PA 16802}
\altaffiltext{5}{DAPNIA/SEDI-SAP, Service d'Astrophysique, CEA-Saclay, 
91191 Gif-sur-Yvette Cedex, France}

\begin{abstract}
We present a catalog of 34 diffuse features identified in X-ray images
of the Galactic center taken with the {\it Chandra} X-ray
Observatory. Several of the features have been discussed in the
literature previously, including 7 that are associated with a complex
of molecular clouds that exhibits fluorescent line emission, 4 that
are superimposed on the supernova remnant Sgr A East, 2 that are
coincident with radio features that are thought to be the shell of
another supernova remnant, and one that is thought to be a pulsar wind
nebula only a few arcseconds in projection from \sgrastar. However,
this leaves 20 features that have not been reported previously. Based
on the weakness of iron emission in their spectra, we propose that
most of them are non-thermal. 
One long, narrow feature points toward \sgrastar, 
and so we propose that this feature is a jet of
synchrotron-emitting particles ejected from the supermassive black
hole. For the others, we show that their sizes (0.1--2 pc in
length for $D$=8 kpc), X-ray luminosities (between $10^{32}$ and
$10^{34}$ \ergsec, 2--8~keV), and spectra (power laws with $\Gamma = 1
- 3$) are consistent with those of pulsar wind nebulae. Based on the
star formation rate at the Galactic center, we expect that $\sim$20
pulsars have formed in the last 300 kyr, and could be producing pulsar
wind nebulae. Only one of the 19 candidate pulsar wind nebulae is
securely detected in an archival radio image of the Galactic center;
the remainder have upper limits corresponding to $L_{\rm
R}$$\la$$10^{31}$ \ergsec. These radio limits do not strongly
constrain their natures, which underscores the need for further
multi-wavelength studies of this unprecedented sample of Galactic
X-ray emitting structures.
\end{abstract}

\keywords{Galaxy: center --- stars: neutron --- supernova remants --- acceleration of particles --- X-rays: ISM}

\section{Introduction}

The Galactic center is host to a high concentration of energetic 
processes, driven by the interplay between the energetic output of several
young star clusters, radiation and outflows powered by accretion 
onto the supermassive black hole \sgrastar, and dense molecular clouds
that have large turbulent velocities. The young stellar
population naturally explains the presence of the mixed-morphology
supernova remnant Sgr A East \citep{mae02,sak04,par05}, a shell-like
radio supernova remnant that exhibits two regions of bright, non-thermal
X-ray emission \citep{ho85,lwl03,sak03,yz05}, and at least two 
X-ray features that resemble pulsar wind nebula 
\citep[e.g.,][Wang, Lu, \& Gotthelf 2006]{par04}.
These processes 
also produce astronomical phenomena unique to the Galactic center.
For instance, fluorescent iron emission is produced where molecular clouds 
are bombarded by hard X-rays from transient sources \citep[most likely 
\sgrastar;][]{smp93,koy96,mkm01,mur01b,par04,rev04,mun07}.
Finally, numerous milliGauss-strength bundles of magnetic fields 
illuminated by energetic electrons produce radio 
\citep[][Yusef-Zadeh, Hewitt, \& Cotton 2004]{yzm87,nor04}\nocite{yzhc04}
and sometimes X-ray emission 
(Wang, Lu, \& Lang 2002; Lu, Wang, \& Lang 2003)\nocite{wll02b,lwl03}; 
the origin of these non-thermal filaments is under debate.

The X-ray emission from the brightest of these features has been described 
by various authors in the references above. 
However, whereas at radio wavelengths comprehensive catalogs of
HII regions and non-thermal features near the Galactic center are available
\citep[e.g.,][]{ho85,nor04,yzhc04},
no similar catalog has been produced in the X-ray band. Here, we remedy 
this by presenting a catalog of X-ray features identified in 
images produced from 1 Msec of 
\chandra\ observations of the central 20 pc of the Galaxy. We focus on
features that are clearly larger than the \chandra\ point spread function,
but that are smaller than $\approx$1.5\arcmin; i.e., we do not include
in our catalog the bulk of the emission from Sgr A East \citep{mae02}, 
or the apparent outflow oriented perpendicular to the Galactic plane
and centered on the Sgr A complex \citep[][and in prep.]{mor03}.

\begin{deluxetable*}{lccccc}[htp]
\tablecolumns{6}
\tablewidth{0pc}
\tablecaption{Observations of the Inner 20 pc of the Galaxy\label{tab:obs}}
\tablehead{
\colhead{} & \colhead{} & \colhead{} & 
\multicolumn{2}{c}{Aim Point} & \colhead{} \\
\colhead{Start Time} & \colhead{Sequence} & \colhead{Exposure} & 
\colhead{RA} & \colhead{DEC} & \colhead{Roll} \\
\colhead{(UT)} & \colhead{} & \colhead{(s)} 
& \multicolumn{2}{c}{(degrees J2000)} & \colhead{(degrees)}
} 
\startdata
2000 Oct 26 18:15:11 & 1561 & 35,705 & 266.41344 & $-$29.0128 & 265 \\
2001 Jul 14 01:51:10 & 1561 & 13,504 & 266.41344 & $-$29.0128 & 265 \\
2001 Jul 17 14:25:48 & 2284 & 10,625 & 266.40417 & --28.9409 & 284 \\
2002 Feb 19 14:27:32 & 2951  & 12,370 & 266.41867 & $-$29.0033 & 91 \\
2002 Mar 23 12:25:04 & 2952  & 11,859 & 266.41897 & $-$29.0034 & 88 \\
2002 Apr 19 10:39:01 & 2953  & 11,632 & 266.41923 & $-$29.0034 & 85 \\
2002 May 07 09:25:07 & 2954  & 12,455 & 266.41938 & $-$29.0037 & 82 \\
2002 May 22 22:59:15 & 2943  & 34,651 & 266.41991 & $-$29.0041 & 76 \\
2002 May 24 11:50:13 & 3663  & 37,959 & 266.41993 & $-$29.0041 & 76 \\
2002 May 25 15:16:03 & 3392  & 166,690 & 266.41992 & $-$29.0041 & 76 \\
2002 May 28 05:34:44 & 3393  & 158,026 & 266.41992 & $-$29.0041 & 76 \\
2002 Jun 03 01:24:37 & 3665  & 89,928 & 266.41992 & $-$29.0041 & 76 \\
2003 Jun 19 18:28:55 & 3549 & 24,791 & 266.42092 & --29.0105 & 347 \\
2004 Jul 05 22:33:11 & 4683 & 49,524 & 266.41605 & --29.0124 & 286 \\
2004 Jul 06 22:29:57 & 4684 & 49,527 & 266.41597 & --29.0124 & 285 \\
2004 Aug 28 12:03:59 & 5630 &  5,106 & 266.41477 & --29.0121 & 271 \\
2005 Feb 27 06:26:04 & 6113 &  4,855 & 266.41870 & --29.0035 & 91 \\
2005 Jul 24 19:58:27 & 5950 & 48,533 & 266.41520 & --29.0122 & 277 \\
2005 Jul 27 19:08:16 & 5951 & 44,586 & 266.41514 & --29.0122 & 276 \\
2005 Jul 29 19:51:11 & 5952 & 43,125 & 266.41509 & --29.0122 & 275 \\
2005 Jul 30 19:51:11 & 5953 & 45,360 & 266.41508 & --29.0122 & 275 \\
2005 Aug 01 19:54:13 & 5954 & 18,069 & 266.41505 & --29.0122 & 275
\enddata
\end{deluxetable*}

\section{Observations}

The \chandra\ X-ray Observatory has observed the inner $\approx$10\arcmin\ 
of the Galaxy with the Advanced CCD Imaging Spectrometer 
imaging array \citep[ACIS-I;][]{wei02} for a total of 1 Msec between 
2000 and 2006 \citep[Table~\ref{tab:obs};][]{bag03,m-cat}.\footnote{We 
omitted the first observation taken on 1999 September 21 from 
our data analysis, because it was taken with the detector at a different
temperature than the remaining observations, and the response of the 
detector is not as well-calibrated at that temperature.} 
The observations were processed using CIAO version 3.3.0 and the 
calibration database version 3.2.2. 
We processed the event lists for each observation by correcting the pulse 
heights of the events for position-dependent charge-transfer inefficiency, 
excluding events that did not pass the standard ASCA grade 
filters and \chandra\ X-ray center (CXC) good-time filters, and removing 
5.7 ks of exposure during which the background rate flared to 
$\ge 3\sigma$ above the mean level. We then refined the default estimate 
of the detector 
positions of each photon received by applying the single-event repositioning
algorithm of \citet{li04}. Finally, we applied a correction to 
the absolute astrometry of each pointing in two steps. First, for the 
deepest observation, we aligned the positions of 23 foreground 
X-ray sources detected at energies below 2~keV within 5\arcmin\
of the aim-point to sources 
in the 2MASS catalog \citep{skr06}. Second, we registered the remaining 
observations to the deepest one by comparing the locations of X-ray 
sources in each image. The resulting astrometric frame should be accurate
to 0\farcs2. A composite image of the field is displayed with 2\arcsec\
resolution in Figure~\ref{fig:image}.

\subsection{Identification of Diffuse Features\label{sec:id}}

We searched for diffuse features using two independently-developed
wavelet-based algorithms. For the first, we used the standard CIAO
tool {\tt wavdetect} \citep{free02} to search for sources, and
identified candidate diffuse features from the resulting list by
searching the raw images (no images were smoothed for this work) by
eye for sources with sizes larger than that of the point spread
function. We ran our search using nine composite images, composed of
three energy bands (the full 0.5--8.0~keV band, the 0.5--2.0~keV band
to increase our sensitivity to foreground sources, and the 4--8~keV
band to increase our sensitivity to highly absorbed sources) and three
spatial resolutions (primarily for computational efficiency: the full
0\farcs5 resolution for the inner 1024$\times$1024 pixels, 1\arcsec\
resolution covering the inner 2048$\times$2048 detector pixels, and
2\arcsec\ resolution covering the entire image).  For the purposes of
source detection only, we removed events that had been flagged as
possible cosmic-ray afterglows. We employed the default ``Mexican
Hat'' wavelet, and used a sensitivity threshold that corresponded to a
$10^{-7}$ chance of detecting a spurious source per PSF element if the
local background is spatially uniform. We expect roughly one spurious
source in our field.  We used wavelet scales that increased by a
factor of $\sqrt{2}$: 1--4 pixels for the 0\farcs5 image, 1--8 pixels for the
1\arcsec\ image, and 1--16 pixels for the 2\arcsec\ image.  The final source
list was the union of the lists from the three images taken
at different pixel scales and energies.
Sources were identified as duplicates
if the they fell within the average radius of the 90\% encircled-energy 
contour for the point spread function (PSF) for 4.5~keV photons at 
the position of the source. When duplicate sources were found in the 
list, we retained the positions from the list derived from the 
highest-resolution images, either the finest pixel scale or the lowest 
energy band. 
Among the list of 2933 sources identified with {\tt wavdetect}, we selected
73 candidate diffuse features. 

\begin{figure*}[htp]
\centerline{\psfig{file=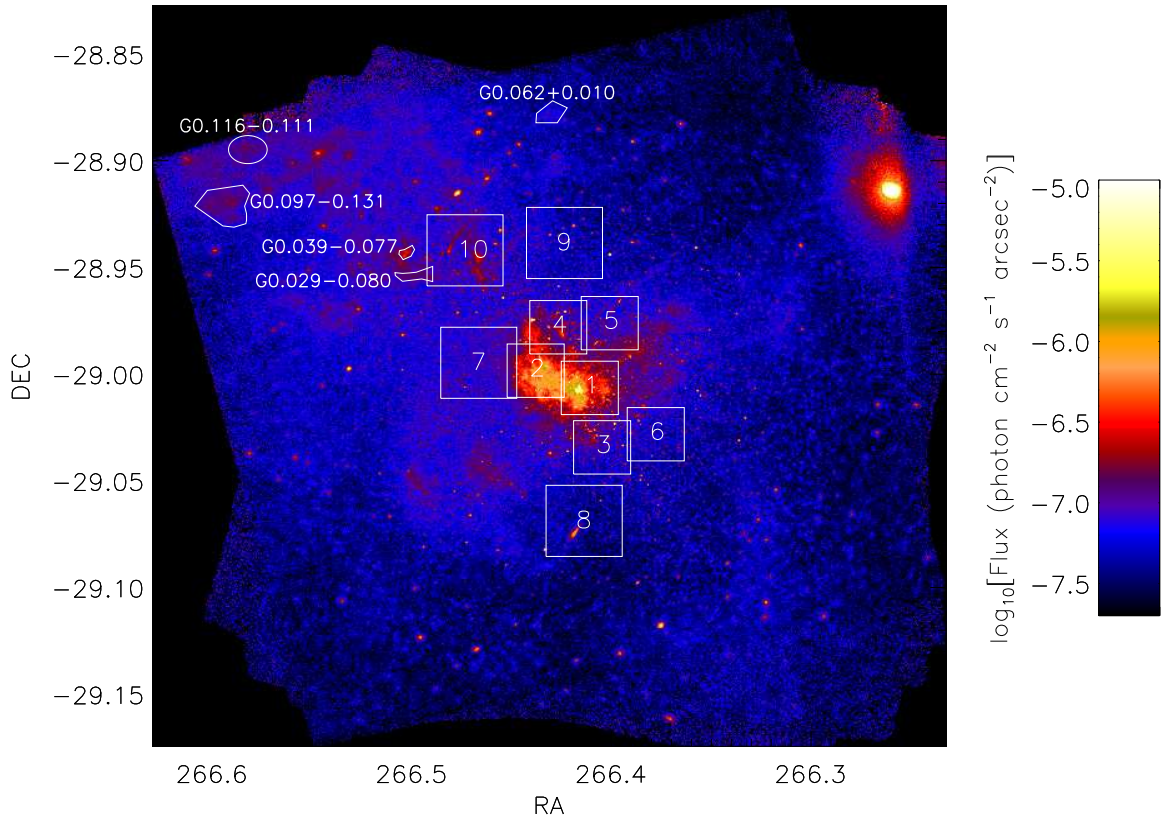,width=0.9\linewidth}}
\caption{
\chandra\ image of the entire survey region, binned to 2\arcsec\
resolution. The numbered boxes denote sub-images that are displayed 
in Figures 2 and 3. 
}
\label{fig:image}
\end{figure*}

\begin{figure*}[htp]
\centerline{\psfig{file=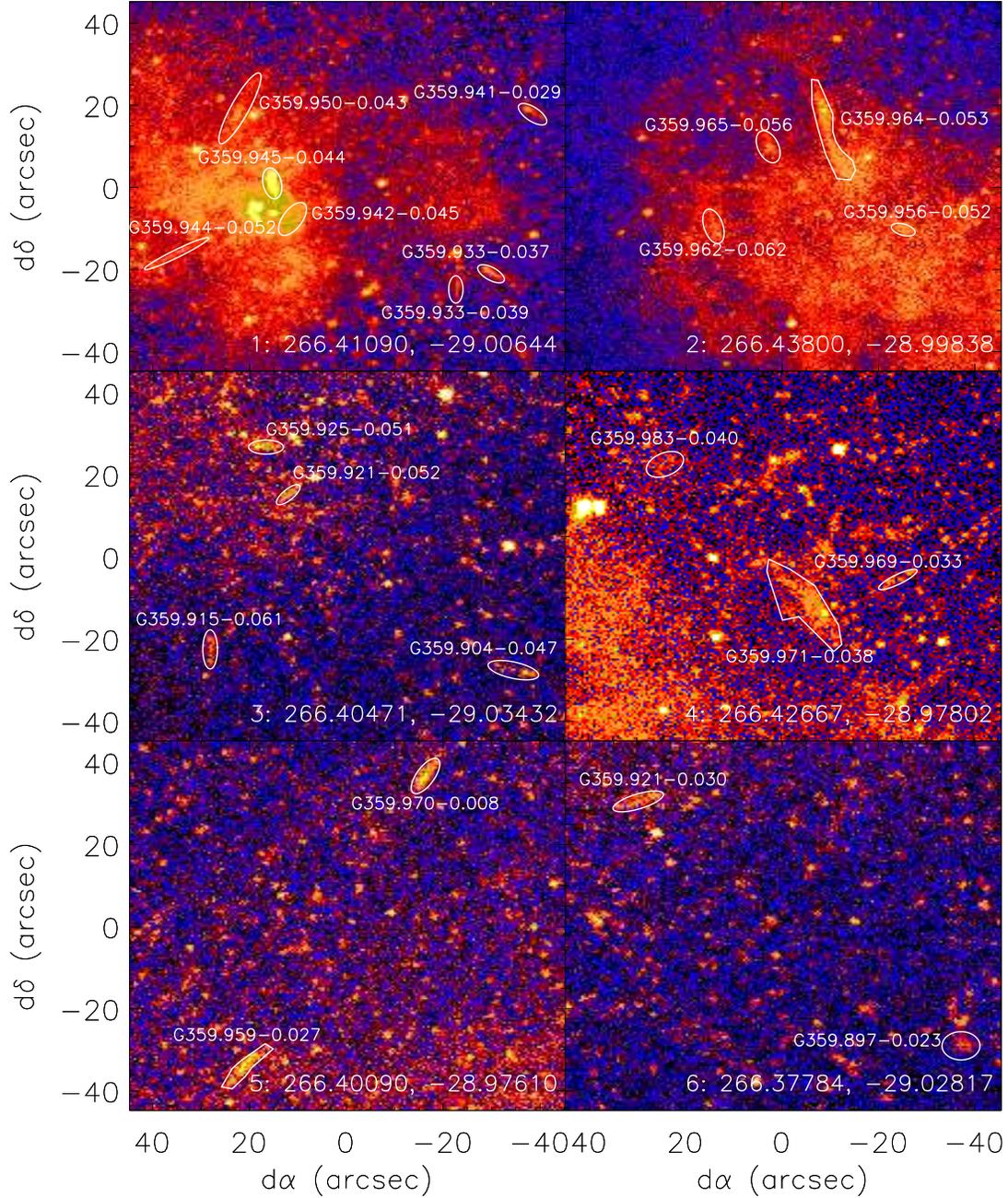,width=0.9\linewidth}}
\caption{
\chandra\ images of regions near the aim-point of the survey that contain 
diffuse features. The coordinates of the center of each image and
its number from Figure~1 are printed at the bottom of each image.
The images are displayed at the full 0\farcs5 resolution of the detector
and telescope. No smoothing has been applied.
Diffuse features that are included in our catalog are outlined in white.
}
\label{fig:sub_small}
\end{figure*}

\begin{figure*}[htp]
\centerline{\psfig{file=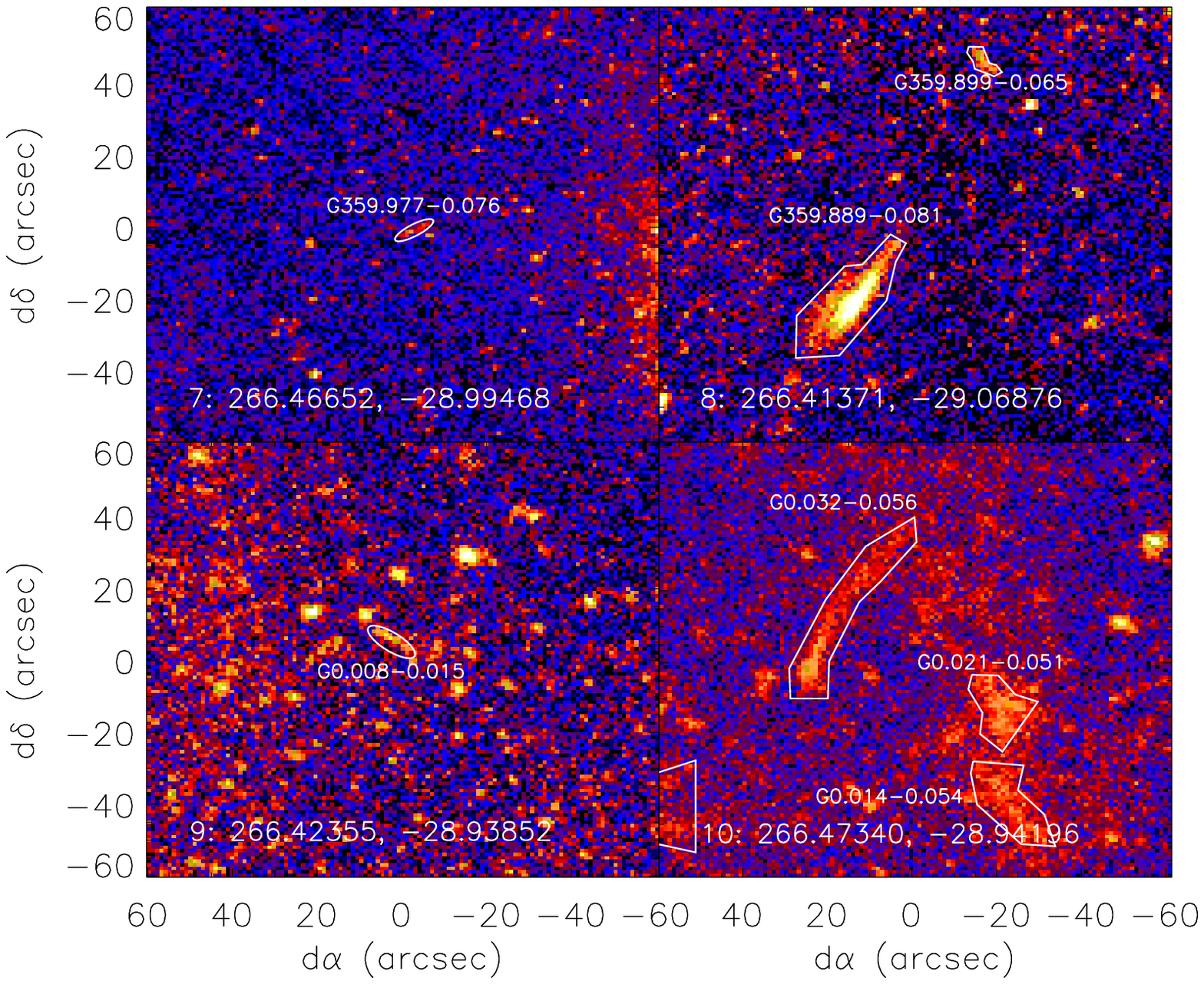,width=0.9\linewidth}}
\caption{
\chandra\ images of regions near $\approx$5\arcmin\ from the center of 
the survey that contain diffuse features, as in Figure 2.
The images are binned to 1\arcsec\ resolution.
Diffuse features that are included in our catalog are outlined in white.
}
\label{fig:sub_med}
\end{figure*}

For the second search, we used the wavelet reconstruction algorithm 
of \citet{sta00}. The algorithm searched for features in the 
0.5--8.0~keV band image that were significant at the $\ge$3$\sigma$ level. 
The results of the decomposition were used to
create an image of the field, and then a multi-scale vision model was
used to extract the structural parameters of 
both point-like and extended sources. We defined as extended those 
sources that either (1) were 
clearly anisotropic, for which the ratio of the major to minor axes
were greater than 1.8, or (2) appeared extended, in that their major
and minor dimensions were at least 3$\sigma$ larger than the mean 
values at that offset from \sgrastar. In total, the algorithm tabulated
2709 objects, of which we identified 91 as anisotropic and 80
as unusually large, for a total of 171 extended sources.

We then compared the lists of candidate sources, and found that only 
36 objects were identified using both wavelet analyses. All of the 
sources that were only in one of the two lists were faint, and visual 
inspection suggested that many were likely to be close alignments of point 
sources. We examined this hypothesis by extracting the average spectra of 
these faint features (the technique is described in \S2.2). We found that the 
average spectra could be described as a $\Gamma = 0.6$ power law absorbed
by a column density of $N_{\rm H} = 1\times10^{23}$ cm$^{-2}$, along with
line emission at 6.7~keV from He-like Fe with an equivalent width of 220 
eV. This spectrum is similar to those of faint point-like sources
in \citet{m-ps}, which supports our hypothesis that these are not truly 
extended features. 

At the same time, at least one feature previously identified as extended is 
not identified as such by either algorithm --- CXOGC 174545.5--285829,
the candidate neutron star suggested to be associated with Sgr A East
by \citet{par05}. Our difficulty in automatically identifying this source 
as diffuse probably stems from the faintness of the extended portion of its
emission. Therefore, we caution that our algorithm for 
identifying extended features is certainly (1) not complete, in the sense 
that it could miss extended features, and (2) contains spurious features
that are chance alignments of point sources. 

We have mitigated the second problem by examining only the 
\numsrc\ sources that are found through both wavelet-based algorithms. The 
remaining 172 objects
will be flagged as being ``possibly extended'' in an updated catalog of 
the point-like sources that we are preparing. In 
Figures~\ref{fig:image}--\ref{fig:sub_med}, we display images with the 
\numsrc\ diffuse features marked. In Figure~\ref{fig:image}, we display the 
entire field, but for readability we only mark the locations of features 
with offsets of at least 6\arcmin\ from \sgrastar, 
and plot boxes around portions of the field closer to the aim point 
that contain diffuse features. These sub-regions are displayed in 
Figures~\ref{fig:sub_small} and \ref{fig:sub_med}. The clustering of sources
in the center of the image results mostly from the fact that the 
\chandra\ PSF is smallest near the aim point, which provides us 
the ability to identify $\approx$1\arcsec\ features as extended.

\subsection{Photometry and Spectroscopy}

In order to compute the number of X-ray events received
from each source, we drew by hand polygonal or elliptical regions 
enclosing each of the candidate diffuse features. The regions are 
plotted in Figures~\ref{fig:image}--\ref{fig:sub_med}. 
We estimated the background
counts in each region using surrounding annular regions that contained 
$\approx$1000 total counts, after excluding both the diffuse 
features and circles enclosing 92\% of photons from each point source
(regions excluding larger fractions of the PSF would cover the 
inner part of the image, leaving no events suitable for estimating a 
local background). We then computed
approximate photon fluxes by dividing the net counts received in 
each of four energy bands --- 0.5--2.0, 2.0--3.3, 3.3--4.7, and 4.7--8.0~keV 
---  by the mean effective area derived using the CIAO tool {\tt mkacisarf}.
These energy bands were chosen to contain roughly equal numbers of counts,
and so that the effective area in each energy range was roughly constant
\citep{m-cat}. In Table~\ref{tab:sources} we report 
the mean location, net counts, and estimated photon fluxes for each source.

\begin{figure*}[htp]
\centerline{\psfig{file=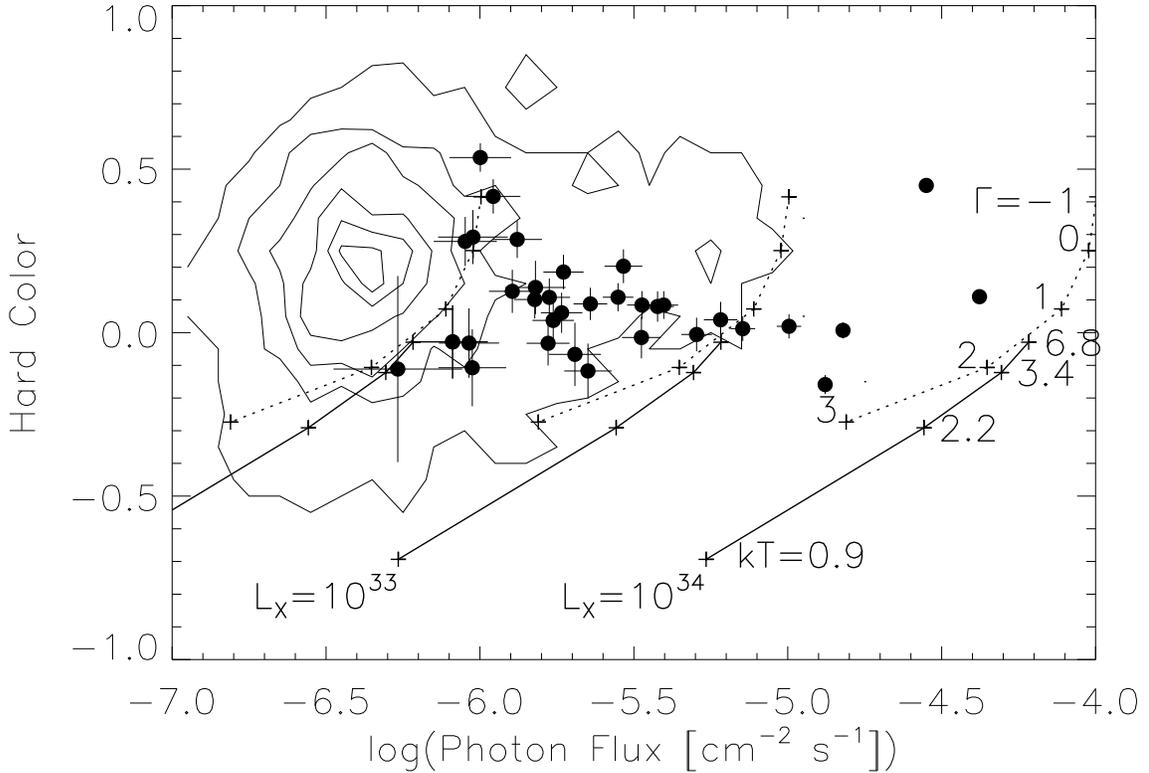,width=0.9\linewidth}}
\caption{
The color-intensity diagram for the diffuse features. The hard color
is defined as $(h-s)/(h+s)$, where $h$ is the number of counts in the
4.7--8.0 keV band, and $s$ is the number of counts in the 3.3--4.7~keV
band.  The solid lines illustrate the hard colors and photon fluxes
expected for thermal plasma with a range of temperatures (as marked on
the right) and luminosities ($L_{\rm X}$=$10^{32}$, $10^{33}$, and
$10^{34}$ \ergsec\ from left to right, for $D=8$~kpc in the
0.5--8.0~keV band), and the dotted lines are for power laws with a
range of photon indices and luminosities. The contours drawn with the
thin lines schematically illustrate the distribution of point-like
X-ray sources in the color-magnitude diagram \citep{m-ps}.  Diffuse
features fainter than $\approx$$2\times10^{-6}$ \phcms\ have mean
colors that are similar to those of the point sources, which suggests
that they could be chance alignments of point sources. However, the
mean color of brighter diffuse features (0.1) is lower than that of
the point-like sources (0.2), which indicates they are a distinct
population.
\label{fig:hit}
}
\end{figure*}

To estimate the intrinsic slopes of the spectra and the absorption toward each 
source, we computed two hardness ratios. Following \citet{m-cat}, the 
ratios were defined as the fractional difference between two energy bands,
$(h-s)/(h+s)$, where $h$ is the number of counts in the hard band, and
$s$ is the number of counts in the soft. The soft color was calculated
using $h$ as the 2.0--3.3~keV band and $s$ as the 0.5--2.0~keV band, and 
was most sensitive to the absorption toward each source. The hard color
was calculated using $h$ as the 4.7--8.0~keV band, and $s$ as the 
3.3--4.7~keV band, and provided a good measure of the intrinsic hardness
of a source \citep{m-ps}. We list the hardness ratios in 
Table~\ref{tab:sources}, and plot them as a function of the 0.5--8.0~keV
photon fluxes in Figure~\ref{fig:hit}. We also plot the hardness
ratios and fluxes that we would expect for a variety of spectral shapes and
luminosities and a single absorption column of $6\times10^{22}$ cm$^{2}$ of
gas and dust, using the thick black lines. 

We find that all but one of diffuse sources have soft colors $>$$-0.175$,
which implies that they are absorbed by $N_{\rm H} > 4\times10^{22}$ cm$^{-2}$
of interstellar medium. This suggests that these sources lie near or 
beyond the Galactic center. One source, G359.977--0.076, has a soft 
color of only
$-0.4^{+0.3}_{-0.4}$, which implies that it is less absorbed, and 
probably lies within a few kpc of Earth. The luminosities of these
sources, assuming they lie at $D$$=$8 kpc, are between $10^{32}$ and
$10^{34}$ \ergsec, and in general their hard colors imply that their spectra
are consistent with either a power law with photon index $\Gamma \approx 1-2$, 
or a $kT$$>$3~keV thermal plasma. For comparison, in Figure~\ref{fig:hit} 
we have also plotted contours denoting the locations of the point-like
sources in this diagram \citep{m-ps}. We find that the diffuse sources are 
slightly softer, with median hard color of 0.1, compared to 0.2 for
the point sources.

\begin{figure}
\centerline{\psfig{file=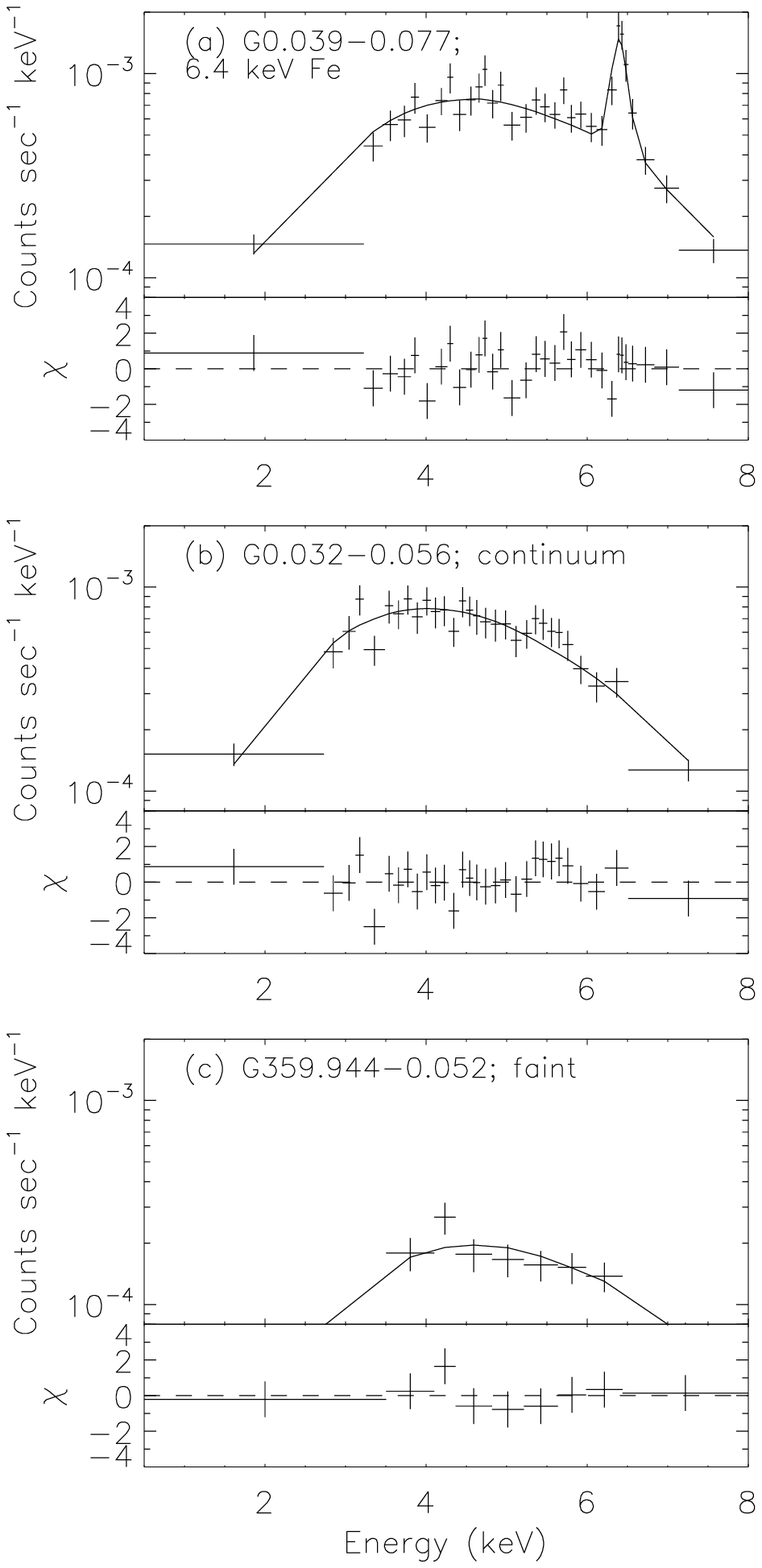,width=0.9\linewidth}}
\caption{
Example spectra of diffuse features. The top panels contain the 
spectra in units of detector counts per second (crosses) and the
best-fit model convolved with the telescope and instrument response
(solid line). The bottom panels contains the residuals after subtracting
the best-fit models from the data, in units of the statistical uncertainty
on the observed spectra. The spectra are modeled with power-law 
continua and Gaussian lines (if necessary), both of which are absorbed 
interstellar gas and dust. Four examples are given: (a) a 
feature that contains a strong 6.4~keV line from low-ionization Fe, 
(b) a bright non-thermal feature lacking line emission,
and (c) a faint source for which there is not a good constraint on the
presence or absence of line emission.
}
\label{fig:spec}
\end{figure}

We then modeled the spectra of the brightest sources in more detail.
Spectra were created by
binning the source and background events over their pulse-heights. The
background-subtracted spectra were then grouped such that each bin had 
a signal-to-noise
ratio of at least 5. We found that because of their extended nature and 
the high background in the image, sources required $\approx$400 net counts
in order to provide 4 independent spectral bins, which is required for us
to constrain independently the interstellar absorption and a two-parameter
continuum model.
For those sources with enough signal, the observed spectra were compared 
with models of an absorbed power law using the chi-squared minimization 
implemented in {\tt XSPEC} version 12.2.0 \citep{arn96}. 
The efficiency and energy resolution of the detector were characterized 
using the above effective-area function and the the energy response 
computed using {\tt mkacisrmf}. 
Representative spectra of 
four sources are displayed in Figure~\ref{fig:spec}. 
Most of the sources could be adequately described by these simple 
absorbed continuum models, but this was largely a result of the low-signal
to noise. The spectral parameters are consistent with those that would
be derived from Figure~\ref{fig:hit}.

However, the models did not adequately describe several of the sources
with $\ga$1000 net counts, and visual inspection revealed that these
exhibited strong emission from low-ionization Fe at 6.4~keV \citep[see
also][]{par04}.  Therefore, for all sources for which the spectra
contained at least 5 independent spectral bins, 
we attempted to model the spectra as the sum of a power law
plus Gaussian line emission at 6.4~keV, absorbed by the
interstellar medium. If the addition of a line to the model resulted
in a decrease in reduced chi-squared, we tested the significance of
the added line using Markov chain Monte Carlo simulations of an $F$
test, as described in Arabadjis, Bautz, \& Arabadjis
(2004)\nocite{aba04} and \citet{m-ps}. The $F$ value is given by
\begin{equation}
F = \frac{(\chi_s^2 - \chi_c^2)}{(\nu_s - \nu_c)} \frac{\nu_s}{\chi_s^2},
\end{equation}
where $\chi_s^2$ and $\nu_s$ are the values of chi-squared and the number of
degrees of freedom for the model without the line, and $\chi_c^2$ and
$\nu_c$ are the same values for the model with a line. We simulated 
a set of 100--1000 spectra with parameters that were consistent with the 
measured parameters of the source (the Markov chain), and computed a 
theoretical distribution of the $F$ value under the assumption that no 
line was present. If the observed value of $F$ exceeded the theoretical
distribution under the null hypothesis in all of 100--1000 trials, 
we considered a line to be detected with $>$99\%--99.9\% confidence. 
In Table~\ref{tab:spec}, we list the parameters 
of the best-fit absorbed power law for each source we modeled, and either 
the measured value of the equivalent width of the 6.4~keV iron line or 
an upper limit to that value. We performed a similar search for the 
6.7~keV line from the He-like 2--1 transition of Fe, and only found one 
source with such a line, G359.942--0.045 (source 2 in Table~\ref{tab:sources}).
The upper limits to the equivalent widths of the
6.7~keV lines from the other sources were generally 25\% higher than those 
on the 6.4~keV lines, for the cases in which neither line was detected.

\begin{deluxetable*}{lcccccccc}[htp]
\tabletypesize{\scriptsize}
\tablecolumns{9}
\tablewidth{0pc}
\tablecaption{Spectral Parameters of the Diffuse Features\label{tab:spec}}
\tablehead{
\colhead{} & \colhead{Object} & \colhead{$N_{\rm H}$} & \colhead{$\Gamma$} & \colhead{Norm} & \colhead{$EW_{\rm Fe}$} & \colhead{$\chi^2/\nu$} & \colhead{$F_{\rm X}$} & \colhead{$L_{\rm X}$} \\
\colhead{} & \colhead{} & \colhead{} & \colhead{} & \colhead{} & \colhead{} & \colhead{} & \colhead{($10^{-14}$)} & \colhead{($10^{32}$)} \\
\colhead{} & \colhead{} & \colhead{($10^{22}$ cm$^{-2}$)} & \colhead{} & \colhead{} & \colhead{(eV)} & \colhead{} & \colhead{(erg cm$^{-2}$ s$^{-1}$)} & \colhead{(erg s$^{-1}$)}
} 
\startdata
1 & G359.945$-$0.044 & $15.3_{- 0.7}^{+ 0.5}$ & $1.74_{-0.13}^{+0.10}$ & $2.7_{-0.6}^{+0.5}\times10^{-4}$ & $<$60 & 187.4/175 & 33 & 66 \\
2 & G359.942$-$0.045 & $ 9.3_{- 0.6}^{+ 0.7}$ & $2.46_{-0.16}^{+0.19}$ & $1.7_{-0.5}^{+0.7}\times10^{-4}$ & $<$220 & 102.6/57 & 9 & 16 \\
3 & G359.944$-$0.052 & $ 4.7_{- 2.1}^{+ 2.0}$ & $-0.17_{-0.33}^{+0.26}$ & $5.0_{-3.2}^{+8.4}\times10^{-7}$ & $<$470 & 4.3/6 & 3 & 2 \\
4 & G359.950$-$0.043 & $ 5.8_{- 0.9}^{+ 2.3}$ & $0.91_{-0.18}^{+0.40}$ & $5.2_{-1.9}^{+9.4}\times10^{-6}$ & $<$140 & 10.2/12 & 4 & 4 \\
6 & G359.956$-$0.052 & $ 3.3_{- 3.2}^{+ 4.8}$ & $-0.39_{-0.47}^{+0.95}$ & $1.7_{-1.4}^{+9.0}\times10^{-7}$ & \nodata & 1.0/1 & 1 & 1 \\
7 & G359.933$-$0.037 & $14.2_{- 2.0}^{+ 5.9}$ & $1.59_{-0.29}^{+0.77}$ & $1.1_{-0.6}^{+5.2}\times10^{-5}$ & $<170$ & 6.7/9 & 2 & 3 \\
8 & G359.941$-$0.029 & $ 8.5_{- 2.7}^{+ 3.4}$ & $0.44_{-0.36}^{+0.57}$ & $1.0_{-0.7}^{+3.0}\times10^{-6}$ & $<$180 & 6.6/5 & 2 & 2 \\
9 & G359.925$-$0.051 & $14.6_{- 2.9}^{+ 4.0}$ & $1.77_{-0.39}^{+0.88}$ & $1.2_{-0.8}^{+5.2}\times10^{-5}$ & $<$2060 & 5.6/5 & 2 & 3 \\ 
10 & G359.964$-$0.053 & $ 9.9_{- 0.8}^{+ 0.9}$ & $1.64_{-0.15}^{+0.20}$ & $6.1_{-1.8}^{+2.7}\times10^{-5}$ & $<$70 & 43.0/53 & 11 & 17 \\
11 & G359.965$-$0.056 &  $ 9.3_{- 3.1}^{+ 4.5}$ & $1.70_{-0.41}^{+0.75}$ & $1.0_{-0.7}^{+4.2}\times10^{-5}$ & \nodata & 6.4/3 & 2 & 3 \\ 
13 & G359.962$-$0.062 & $ 4.2_{- 1.6}^{+ 3.6}$ & $0.64_{-0.39}^{+0.36}$ & $1.2_{-0.8}^{+7.1}\times10^{-6}$ & \nodata & 6.6/2 & 2 & 2 \\
14 & G359.959$-$0.027 & $16.5_{- 2.0}^{+ 3.9}$ & $2.03_{-0.31}^{+0.76}$ & $0.4_{-0.2}^{+1.2}\times10^{-4}$ & $<$75 & 6.5/12 & 3 & 6 \\
15 & G359.971$-$0.038 & $12.6_{- 1.7}^{+ 2.0}$ & $1.74_{-0.25}^{+0.45}$ & $4.0_{-2.0}^{+5.3}\times10^{-5}$ & $<$130 & 32.8/22 & 6 & 10 \\
17 & G359.921$-$0.030 & $17.1_{- 3.7}^{+ 3.6}$ & $2.48_{-0.55}^{+0.87}$ & $3.8_{-3.0}^{+15}\times10^{-5}$ & $<$1300 & 5.0/5 & 1 & 3 \\
20 & G359.904$-$0.047 & 2.2 & --1.1 & $4.4\times10^{-8}$ & \nodata & 4.7/3 & 1 & 1 \\  
22 & G359.970$-$0.008 & $17.8_{- 2.7}^{+ 3.2}$ & $2.22_{-0.36}^{+0.62}$ & $4.3_{-2.7}^{+9.8}\times10^{-5}$ & $<$110  & 5.9/11 & 2 & 6 \\
23 & G359.899$-$0.065 & $26.2_{- 5.1}^{+ 8.8}$ & $1.86_{-0.85}^{+1.05}$ & $1.3_{-1.1}^{+9.4}\times10^{-5}$ & \nodata & 4.4/3 & 1 & 3 \\ 
24 & G359.897$-$0.023 & $12.7_{- 5.6}^{+ 4.4}$ & $2.66_{-0.60}^{+1.06}$ & $2.1_{-1.9}^{+15}\times10^{-5}$ & \nodata & 0.1/1 & 1 & 2 \\
25 & G359.889$-$0.081 & $29.3_{- 1.1}^{+ 1.8}$ & $1.24_{-0.18}^{+0.27}$ & $1.3_{-0.4}^{+1.0}\times10^{-4}$ & $<$30 & 86.9/95 & 26 & 70 \\
26 & G0.014$-$0.054 & $17.3_{- 7.6}^{+ 4.0}$ & $-0.11_{-0.13}^{+0.11}$ & $2.0_{-4.0}^{+0.4}\times10^{-6}$ & $ 866_{- 252}^{+ 300}$ & 11.6/14 & 8 & 11 \\
28 & G0.021$-$0.051 & $15.6_{- 8.5}^{+ 4.4}$ & $-0.51_{-0.13}^{+0.11}$ & $8.4_{-1.8}^{+1.8}\times10^{-7}$ & $ 998_{- 312}^{+ 378}$ & 15.4/11 & 7 & 9 \\
29 &  G0.032$-$0.056 & $ 7.8_{- 1.4}^{+ 1.2}$ & $1.27_{-0.24}^{+0.29}$ & $2.0_{-0.9}^{+1.4}\times10^{-5}$ & $<$110 & 26.6/27 & 8 & 10 \\
31 & G0.039$-$0.077 & $10.4_{- 1.8}^{+ 3.0}$ & $0.43_{-0.12}^{+0.13}$ & $6.3_{-1.0}^{+1.0}\times10^{-6}$ & $ 661_{- 183}^{+ 230}$ & 28.8/28 & 11 & 13 \\
32 & G0.062+0.010 & $ 9.0_{- 1.4}^{+ 5.1}$ & $1.83_{-0.36}^{+0.21}$ & $3.3_{-0.8}^{+1.0}\times10^{-5}$ & $1576_{- 895}^{+ 921}$ & 22.8/11 & 6 & 9 \\
33 & G0.097$-$0.131 & $20.1_{- 3.3}^{+ 2.5}$ & $1.91_{-0.07}^{+0.06}$ & $4.7_{-0.7}^{+0.8}\times10^{-4}$ & $1193_{- 231}^{+ 275}$ & 135.4/50 & 48 & 100 
\enddata
\tablecomments{All uncertainties are 1$\sigma$, determined by varying the 
parameter of interest until $\chi^2$ increased by 1.0. Limits on the equivalent
width of the iron lines were only computed when there were at least 6 
independent spectral bins. The flux and luminosity were computed for the 
2--8~keV band.}
\end{deluxetable*}

\begin{figure}
\centerline{\psfig{file=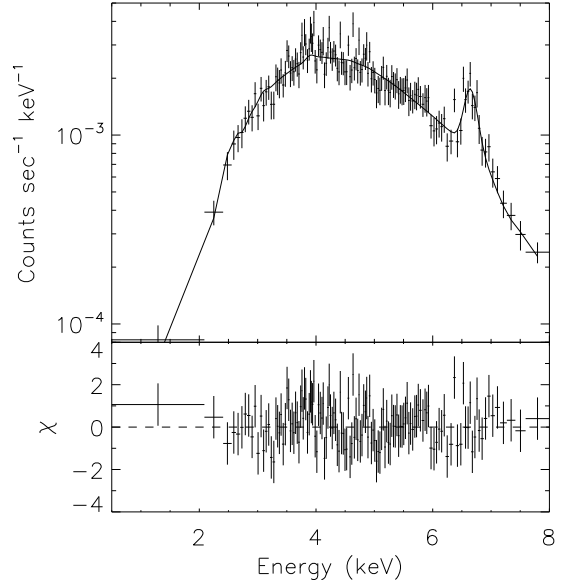,width=0.9\linewidth}}
\caption{
Combined spectra of the 20 faintest diffuse features, 
as in Fig.~\ref{fig:spec}.
Line emission is detected at 6.7~keV, which suggests that there is 
a contribution either from diffuse thermal plasma or point-like 
X-ray sources to these diffuse features. 
}
\label{fig:ave}
\end{figure}

Finally, for the 20 faintest features, we extracted their average spectrum 
(Fig.~\ref{fig:ave}). The background was computed
from an elliptical region enclosing all of the sources, but excluding
the Sgr A complex. The effective area function 
was computed using {\tt mkwarf}, and the response function using
{\tt mkacisrmf}. We found that the spectrum could be modeled 
as a $\Gamma = 1.3^{+0.1}_{-0.2}$ power law absorbed by 
$N_{\rm H} = (1.2\pm0.1) \times 10^{23}$ cm$^{-2}$, along with a 6.7~keV
iron line with an equivalent width of 150$^{+30}_{-70}$ eV 
($\chi^2/\nu = 92.0/88)$. 
The spectrum of the emission from these diffuse features is sightly steeper
than that of the faint point sources (with $\Gamma = 0.9$), and the 
He-like Fe emission is weaker 
\citep[400$\pm$60 eV for point sources][]{m-ps}. This suggests that some, 
but not all, of these faint features are produced by chance alignments 
of point sources. Alternatively, the presence of the He-like Fe line 
could be explained if a diffuse thermal plasma
contributes some of the flux. We therefore also modeled the emission
as a non-equilibrium-ionization plasma, and found a best-fit 
temperature of $kT = 20^{+16}_{-5}$~keV, 
metal abundances of $0.31^{+0.5}_{-0.5}$ solar, 
an ionization time scale 
$\tau = n_e t = 9_{-2}^{+2}\times10^{10}$ s cm$^{-3}$, 
and an absorption column of 
$N_{\rm H} = 1.1^{+0.5}_{-0.6} \times10^{23}$ cm$^{-2}$. 
We examine whether it is reasonable to presume these diffuse features
are composed of hot plasma in \S3.

\subsection{Radio Fluxes}

\begin{figure*}[htp]
\centerline{\psfig{file=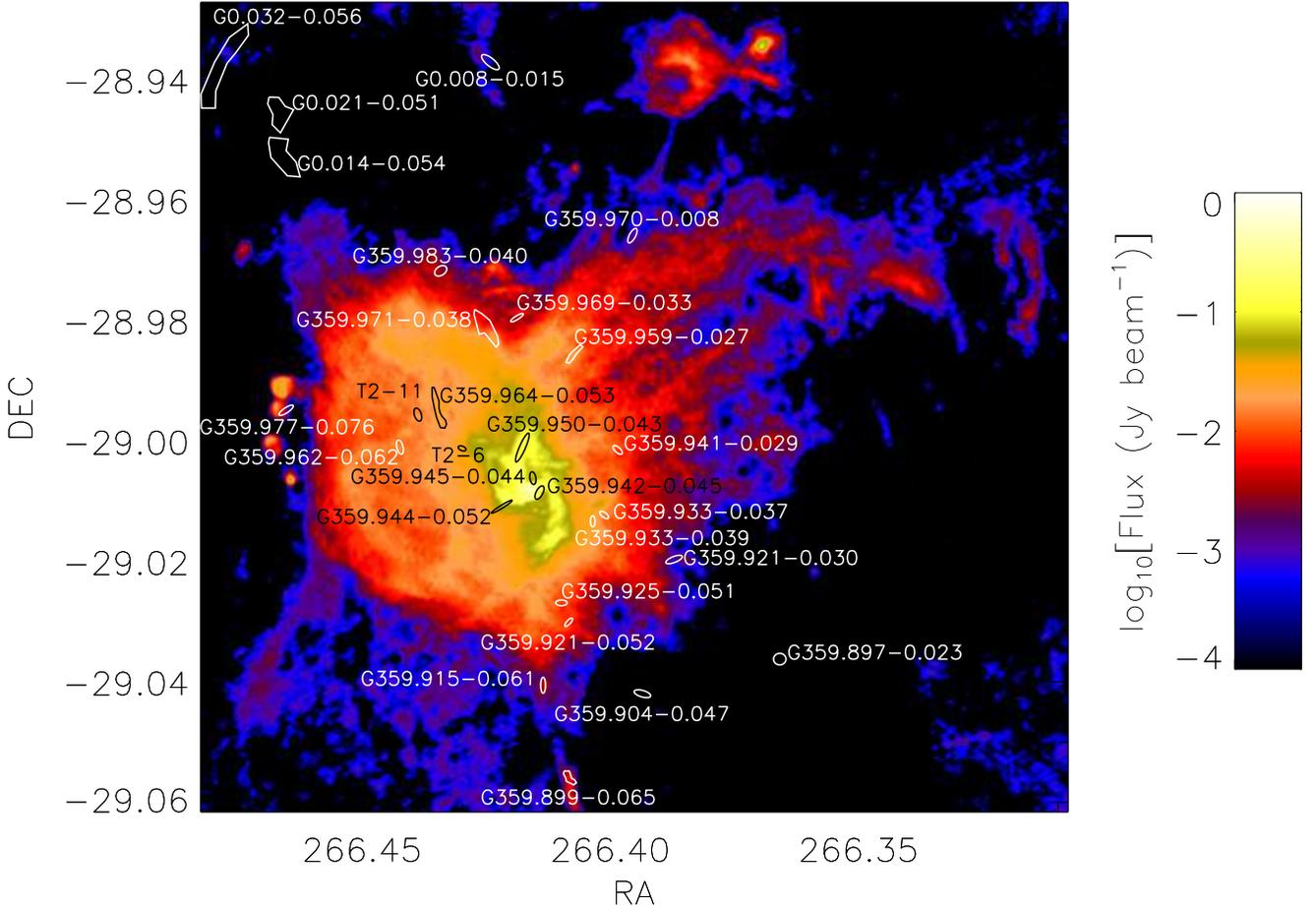,width=\linewidth}}
\caption{
Radio image taken at 6 cm with the VLA \citep{yzm87}, 
covering the central portion
of the \chandra\ image where most of the diffuse X-ray features are found. 
The diffuse X-ray features are marked in either black or white. For 
readability, two sources are marked with their numbers from Table~2: 
G359.956--0.052 (source 6), and G359.965-0.056 (source 11). The
beam size for the radio image is 3\farcs4 by 2\farcs9. 
}
\label{fig:radiob}
\end{figure*}

Several diffuse X-ray features elsewhere in the Galactic center
region have been associated with radio sources \citep{sak03,lwl03,yz05}.
Therefore, we have examined an archival image of the Galactic center
taken at 6 cm with the Very Large Array 
\citep{yzm87}\footnote{Available from 
{\tt http://adil.ncsa.uiuc/QueryPage.html}.} in order to search for
counterparts to the X-ray features. The image is displayed in 
Figure~\ref{fig:radiob}, with the regions from Figures 1--2 overlaid. 
The beam size was 3\farcs4 by 2\farcs9.
To compute fluxes for each feature contained within the radio image, we 
integrated the source flux over each feature and the background flux from 
an annular region around the feature, and subtracted the two. The 
background region had a minimum radius of 5\arcsec\ for the most compact 
features, and was 50\% bigger than the radius of the source region 
for the larger features. As an estimate of the uncertainty in the flux 
measurement, we computed the standard deviation of the flux from the source
region. The fluxes and standard deviations are listed in 
Table~\ref{tab:sources}.

Using the image in Figure~\ref{fig:radiob}, we obtained
upper limits to the radio fluxes for the majority of the sources. 
The limits ranged from a few mJy at an offset of $\ga$1\arcmin\ 
from \sgrastar, to tens of mJy for objects located within the bright 
radio emission of the Sgr A complex. Only three sources have radio fluxes 
measured above the 5$\sigma$ level: G359.950--0.043 (source 4 in 
Tab~\ref{tab:sources}), 
G359.899--0.065 (source 23 in Tab.~\ref{tab:sources}, and
source F in \citealt{yz05}), and G359.959--0.027 (source 14 in 
Tab.~\ref{tab:sources}). Of these, we believe that the association 
between the radio and X-ray emission is physical only for the latter 
two sources.
G359.950--0.043 (source 4) lies on the ``minispiral,'' a feature with 
prominent Hydrogen recombination emission in the optical and radio, and
that is therefore not hot enough to be an X-ray source on its own 
\citep[$kT \sim 1$ eV;][]{sco03}. 

Two regions (G0.021-0.051 and 
G0.032--0.056) with offsets $>$4.5\arcmin\ that fell within the image 
contained negative artifacts left by the cleaning algorithm, so we could
not compute fluxes for them. Better measurements and limits would 
require a dedicated re-analysis of the archival VLA observations, which 
is beyond the scope of this paper.

\section{Discussion\label{sec:dis}}

We can roughly divide the diffuse features in the Galactic center into
two groups, based on their spectral properties and sizes. 
Three of the four features with areas larger than 100 arcsec$^{2}$ exhibit 
strong fluorescent iron emission with equivalent widths $>$500 eV 
(the exception is G359.889-0.081, which we will return to 
shortly). These features are discussed extensively
in \citet{par04} and \citet{mun07}, and are produced because hard X-rays 
($>$7~keV) from a transient source are being scattered by molecular 
gas, and causing iron to fluoresce. These
features all lie in the northeast of the field. In fact, this entire 
quadrant of the image exhibits unusually strong diffuse Fe K-$\alpha$ 
emission \citep[equivalent width 570 eV;][]{m-diff}.
The features that we identify in this paper are simply the brightest 
manifestations of this pervasive scattered and fluorescent emission, 
and they probably trace
regions of molecular gas with the highest optical depth. It is notable, 
however, that similar features are not seen toward other clouds
with similarly high optical depth, particularly those
with velocities of $+$20 km s$^{-1}$ and $+$40 km s$^{-1}$ just to the
east of \sgrastar\ (e.g., G\"{u}sten, Walmsley, \& Pauls 1981; 
Oka \etal\ 1998)\nocite{gwp81,oka98}. 
The clouds that do not exhibit fluorescent emission are
slightly in the foreground of the field, as they are seen to cast a 
shadow in the number of X-ray sources along their lines of sight 
\citep{m-cat}. The clouds without associated X-ray emission must not 
intersect the light-front of the transient
that illuminated the fluorescent features. 

The features with surface area $<$100 arcsec$^2$ either exhibit no 
evidence for iron emission at 6.4~keV, or are too faint
for such emission to be detected.  Based on their spectra, the 
X-ray emission from these features must either originate from a 
hot thermal plasma ($kT \ga 7$~keV; Fig.~\ref{fig:hit}), or from 
non-thermal processes such as synchrotron or inverse-Compton scattering. 
As we describe below, previous work on individual features suggests 
that they have a variety 
of origins. This is the first paper to provide a comprehensive compilation 
of their properties. For brevity, we will refer to the features without 
obvious fluorescent iron emission as ``streaks.'' Below, we speculate about 
their natures.

\subsection{Thermal Shocks}

The possibility that some of the features are thermal is suggested by
the detection of the 6.7~keV line from He-like iron in the combined
spectra of the faint diffuse features (Fig.~\ref{fig:ave}). However, 
this hypothesis requires that some mechanism must be found
to confine the plasma. For a characteristic temperature of $\approx$7~keV,
and a luminosity of $L_{\rm X} = 10^{33}$ \ergsec\ (2--8~keV), the 
emission measure is 
$\int n_e n_{\rm H} dV \approx n_e^2 V = 1\times10^{56}$ cm$^{-3}$, 
where $n_e$ is the electron density, $n_{\rm H}$ is the hydrogen 
density, and $V$ is the volume of the emitting region. The 
sizes of these features are typically $\Omega \approx 20-100$ arcsec$^{2}$,
which for $D$$=$8 kpc corresponds to projected areas of 
$A \approx 0.03-0.15$ pc$^{2}$. If we assume that the volume is given by
$V = A^{3/2}$, then the electron density of the plasma is
$n_e = 13 \Omega_{50}^{-3/4} L_{\rm X,33}^{1/2}$ cm$^{-3}$, where $\Omega_{50}$
is normalized to 50 arcsec$^{2}$, and $L_{\rm X,33}$ to $10^{33}$ \ergsec\
(2--8~keV). Therefore, the pressure is 
$P = 1\times10^{-7} \Omega_{50}^{-3/4} L_{\rm X,33}^{1/2}$ erg cm$^{-3}$.
This pressure is $\approx$300 times larger than that of the diffuse,
$kT \approx 0.8$~keV plasma in the Galactic center 
($3\times10^{-10}$ erg cm$^{-3}$), and 50
times larger than the putative 8~keV plasma there 
\citep[$2\times10^{-9}$ erg cm$^{-3}$;][]{m-diff}.\footnote{The
8 keV plasma may be produced by faint, unresolved point sources 
(Revnivtsev, Vikhlinin, \& Sazonov 2007)\nocite{rvs07}.}
Magnetic fields would have to be strong to confine this plasma,
$B = 2 \Omega_{50}^{-3/8} L_{\rm X,33}^{1/4}$ mG. These fields strengths 
are thought to be present within radio-emitting filaments toward the
Galactic center \citep{ms96}, but the average field at the Galactic center
may be significantly lower \citep[e.g.,][]{lar05}.  Thermal features 
could also be produced by shocks with velocities of 
$\sim$$(4kT/3\mu m_p)^{1/2}$=1200 km s$^{-1}$ (where $m_p$ is the proton mass
and $\mu$=0.5 for a plasma of electrons and protons). However, there are 
no obvious candidates for the sources of the energy powering these 
putative shocks; they are not arranged as part of a shell as would be expected
if they originate from a supernova (and the lack of metal lines would
be puzzling), and the known massive stars in the region are mostly located
within 30\arcsec\ of Sgr A*, which is several parsecs in projection from
these features \citep[e.g.,][]{kra95,pau06}.

\subsection{Supernova Remnants} 

Given the problems with assuming that these features are thermal sources,
we follow other authors and suggest that the majority of these 
features are non-thermal.
A few of the X-ray streaks appear to be non-thermal particles accelerated
in shocks where supernova remnants impact the interstellar medium.
It has been noted by several authors that G359.889--0.081 and G359.899--0.065
(sources 25 and 23 in Table~\ref{tab:sources}) 
lie on part of a shell-like feature seen in radio maps 
\citep{sak03,lwl03,yz05}. These features are coincident with the brightest
parts of the radio-emitting shell, which are referred to as E and F by 
\citet[][and Fig. \ref{fig:radiob}]{ho85}. 
We also find four filamentary features associated with the Sgr A East 
supernova remnant, G359.962--0.062, G359.965--0.056, G359.964--0.053,
and G359.956--0.052 (sources 6, 10, 11, and 13 in Table~\ref{tab:sources}). 
Of these features associated with supernova remnants,
we can put strict upper limits to the equivalent width of iron emission for
G359.956-0.052 and G359.889--0.081, $<$70 and $<$30 eV, respectively, which 
strongly suggests that they are non-thermal.
The X-rays are probably synchrotron emission 
from TeV electrons \citep{yz05}, which could occur in parts of remnants 
where the shock velocity is high and the ambient density is low, so 
that particles are efficiently accelerated \citep[e.g.,][]{vink06}. 
Supernova shocks therefore can explain 5 of the 26 streaks.

\subsection{Non-Thermal Radio Filaments}

At least one non-thermal radio filament has been identified with 
X-ray emission in the central 100 pc of the Galaxy, but outside of our 
deep images of the central 20 pc: G 359.54+0.18 \citep{lwl03}. 
This confirmed non-thermal filament produces a flux density of 
150 mJy at 6 cm over the region that also emits X-rays. Assuming an
$\alpha=-0.8$ spectrum, the estimated radio luminosity of G 359.54+0.18
between $10^{7}$ Hz and $10^{11}$ Hz is $L_{\rm R} = 5\times10^{32}$ \ergsec,
compared to an X-ray luminosity of $L_{\rm X} = 2.4\times10^{33}$ \ergsec\
\citep[0.2--10~keV][]{lwl03}.
In our image, however, only one diffuse feature that is not associated
with a supernova remnant has a firm radio counterpart, G359.959--0.027
(source 14 in Table~\ref{tab:sources}). Its X-ray luminosity is
$L_{\rm X} = 6\times10^{32}$ \ergsec\ (2--8~keV).  In
Fig.~\ref{fig:radiob} \citep[][]{yzm87}, the feature has a flux
density of $20\pm3$ mJy, which for the same assumed radio spectrum as
above implies $L_{\rm R} = 2\times10^{31}$ \ergsec. A second nearby
feature with a similar size, spectrum, and even orientation,
G359.970--0.008 (Figure~\ref{fig:sub_small}, panel 5; source 22 in
Table~\ref{tab:sources}), also has a marginal radio counterpart with a
flux of $1.2\pm0.4$ mJy. For this source we find $L_{\rm X} =
6\times10^{32}$ \ergsec\ (2--8~keV) and $L_{\rm R} \approx
9\times10^{29}$ \ergsec.  Therefore, a couple of the streaks could be
counterparts to non-thermal radio filaments, although the relative
amount of radio and X-ray flux appears highly variable.

\subsection{Pulsar Wind Nebulae}

The final conventional explanation for the X-ray streaks is that
they could be pulsar wind nebulae.
Two candidate pulsar wind nebulae in this field have been 
discussed in the literature. \citet{wlg06} have demonstrated that
the spectral and morphological properties of G359.945--0.044
resemble a pulsar wind nebula, and have suggested that it is the
source of the TeV emission that originates near \sgrastar\ 
\citep{aha04}. This source is included in our catalog. 
\citet{par05} have identified the point-source
CXOGC J174545.5--285829 as another candidate pulsar based on 
its featureless non-thermal spectrum, its lack of long-term variability, 
and the detection of a faint extended tail that may represent a bow 
shock formed by the wind nebula. This object is not in our catalog, 
because the faint nebulosity was not identified by our wavelet algorithms
(see \S\ref{sec:id}). From this new work, one bright feature
has a morphology that makes it an especially attractive candidate
to be a pulsar wind nebulae: G 0.032--0.056 (source 29 in 
Table~\ref{tab:sources}), which has a bright head
producing about half its flux, and a long tail that curves to the 
northwest. 

Pulsar wind nebulae are formed when a wind of particles that is
powered by the rotational energy of the neutron star encounters the
ISM. The resulting shock accelerates the particles further, causing
them to radiate synchrotron emission \citep[see][for a review]{gs06}.
These wind nebulae have diverse morphologies, which depend upon how
the pulsar wind interacts with the surrounding supernova remnant, how
far the pulsar travels during the lifetime of the electrons in the
wind nebula, and whether the pulsar is moving supersonically with
respect to the ISM. In the Galactic center, few of the pulsars are
likely to be moving highly supersonically.  Most of the region is
probably suffused with hot $kT\approx0.8$~keV plasma with densities of
$n\approx0.1$ cm$^{-3}$, which implies that the sound speed of the ISM
is $\approx$500 km s$^{-1}$ \citep{m-diff}.\footnote{We ignore the 8
keV plasma here, for two reasons. First, it would imply a higher
ambient sound speed, and therefore bow-shock nebulae would be even
less likely to form. Secondly, it may be produced by unresolved point
sources \citep{rvs07}.} In contrast, the mean three-dimensional velocities 
of known radio pulsars are $\approx$400 km s$^{-1}$ 
\citep[e.g.,][]{hobbs05, fgk06}, and only 2 of 169 pulsars 
in \citet{hobbs05} have measured transverse velocities
$>$800~km~s$^{-1}$, which, assuming the velocities are isotropic,
would translate to expected three-dimensional velocities
$>$1000~km~s$^{-1}$.  Pulsars will only be highly
supersonic (Mach numbers $>$2) in regions of the ISM that are dense
and cool, such as near molecular clouds.  Therefore, we would not
expect bow-shock nebulae to be common
\citep[although see][]{par05}. Nonetheless, the complex morphologies we 
observe could be explained by interactions between wind nebulae and 
unseen supernova remnants \citep[e.g.,][]{gae03},
or if the proper motion of the pulsar moves the particle acceleration
region faster than the electrons can cool 
\citep[e.g., SN G371.1-1.1 in][]{gs06}. The characteristic 
sizes of known pulsar nebulae are between $10^{17}$ and $10^{19}$ cm
(Cheng, Taam, \& Wang 2004)\nocite{ctw04}, which at the Galactic
center ($D$=8 kpc) would have angular sizes of 0\farcs8--80\arcsec.
This bounds the sizes of the features we observe.

Moreover, the X-ray  luminosities of the diffuse features in
the Galactic center are consistent with those of known pulsar wind
nebulae.  Galactic pulsar wind nebulae have X-ray luminosities that are as
large as $\sim$$10^{37}$
\ergsec\ (2--10~keV); typically, $\sim$$10^{-3}$ of the spin-down 
energy $\dot{E}$ is emitted as X-rays \citep[e.g.,][]{bt97,ctw04,gs06}.
The diffuse features in our survey all have luminosities of 
$L_{\rm X} \sim 10^{32}$--$10^{34}$ \ergsec, which would imply 
spin-down luminosities of $\dot{E} \sim 10^{35} - 10^{37}$ \ergsec.
There are $\approx$86 known pulsars with $\dot{E}$ in this range. 
Most of these energetic pulsars have ages $\la$$10^{5}$ years, although 7 
of $\sim$100 known recycled millisecond pulsars have $\dot{E}$ this high
\citep{man05}.\footnote{Taken from the on-line catalog at 
{\tt http://www.atnf.csiro.au/research/pulsar/psrcat/}}

Many Galactic pulsar wind nebulae are luminous at radio wavelengths, 
and yet we find that most of our X-ray sources do not have radio 
counterparts. The limits to their flux densities at 6 cm are 
$\la$3 mJy (Tab.~\ref{tab:sources} and Fig.~\ref{fig:radiob}).
If we assume a standard pulsar wind spectrum with logarithmic slope
$\alpha = -0.3$ between $10^{7}$ and $10^{11}$ Hz, our typical limit
of 3 mJy corresponds to $L_{\rm R} \sim 10^{31}$ \ergsec, or 
between $10^{-1}$ and $10^{-3}$ of the X-ray luminosity. This would
translate to $\sim (10^{-4}-10^{-6})\dot{E}$.
For comparison, the youngest, most energetic Galactic pulsars 
are detected with $L_{\rm R} \sim 10^{-4} \dot{E}$. However, for 
$\dot{E} \la 10^{36}$ \ergsec\ a few pulsars have 
$L_{\rm R} \la 10^{-5} \dot{E}$ \citep{fs97, gae00}: of 
$\approx$30 pulsars that have been reasonably 
well-studied at X-ray \citep{bt97,ctw04} and radio \citep{fs97,gae00} 
wavelengths, five stand out as having wind nebulae 
with $L_{\rm X} > 10^{32}$ \ergsec\ and 
$L_{\rm R} \la (10^{-5}-10^{-7})\dot{E}$: PSR B1046--58 \citep{gon06}, 
PSR B1823--13 \citep{gae03}, PSR B0355+54 \citep{mcg06}, 
PSR B1706-44 \citep{ghd02,rom05}, and possibly PSR J1105--6107 
\citep[][]{gk98}.
The sources in our sample that are either bright in X-rays or that have
strict limits to their radio fluxes would have to be analogous to 
these radio-faint pulsars. 

Given that the lack of radio detections can be reconciled plausibly with the
known properties of pulsars, we next examine whether
we should expect enough energetic pulsars in the Galactic
center to account for the number of diffuse features that we observe.
\citet{wjc05} and \citet{ctw06} have considered whether recycled millisecond
pulsars in the Galactic center could contribute to these X-ray features. 
Based on the inferred population of Galactic millisecond pulsars 
\citep{lyn98}, they estimate that $\sim$200 could be present in our survey
region. If, based on the catalog of \citet{man05}, $\sim$10\% of the 
millisecond pulsars have $\dot{E} > 10^{35}$ \ergsec, they also could
contribute $\sim$20 objects to the diffuse features at the Galactic center. 
These pulsars will generally be traveling rather slow 
\citep[$v$$\sim$130 km s$^{-1}$][]{lyn98}, relative to the sound speed of
the $\approx$0.8~keV plasma that takes up most of the volume of the ISM at
the Galactic center \citep{m-diff}, but they could form extended nebulae
if they encounter regions of the ISM with relatively low sound 
velocities ($kT$$\la$0.1~keV). The filling factor of this cooler ISM is 
unknown, but if it is $\ga$10\%, a handful of millisecond
pulsars could produce extended diffuse features. 

Most of the known energetic pulsars with wind nebulas are 
young, and still interacting with their supernova remnants. We 
concentrate on these as the best candidates for the diffuse 
X-ray features. 
\citet{fig04} modeled the infrared color-magnitude diagrams of stars
in several fields toward the Galactic center, and, assuming an initial
mass function (IMF) of the form $dN \propto M^{-0.9}$ for stars with
masses between 0.1 \msun\ and 120\msun,\footnote{This IMF is flatter than 
in usually assumed for the Galactic disk \citep[e.g.,][]{krou02}, but we do
not expect this to be a large source of uncertainty, because the models
were generated to reproduce the number of luminous, and therefore
relatively more massive, stars \citep{fig04}.}
concluded that the star
formation rate in the inner 30 pc of the Galaxy is 0.02 \msun\
yr$^{-1}$. 
If we assume that radio pulsars form from stars with initial
masses between 8 and 30 \msun\ \citep[e.g.,][but see Figer \etal\
2005; Muno \etal\ 2006]{bj99,heg03}, then they would form at a rate of
$2\times10^{-4}$ yr$^{-1}$.  

However, the supernovae that produce pulsars provide them with kicks, 
so the loss of pulsars as they move out of the Galactic center will also limit 
the number of pulsar wind nebulae observable there 
\citep[e.g.,][]{cl97}.\footnote{\citet{pl04} discuss millisecond radio 
pulsars that are $10^{8}$ yr old located within 0.02 pc of \sgrastar. These 
have higher escape velocities and would be retained, but they would be
faint and confused with \sgrastar\, and so would not be detected in 
our X-ray survey.}  
There is still debate as to the three-dimensional velocity distribution
of pulsars \citep[e.g.,][]{arz02,hobbs05,fgk06}, so we have adopted 
a simplified approach to estimate the number of pulsars remaining
within the 8\arcmin\ (20 pc) radial bound of our image, based upon 
the observed tangential velocities for 169 radio pulsars that were 
tabulated by \citet{hobbs05}. 
We assume that a pulsar would remain in our image if  the product of 
its tangential 
velocity $v$ and its lifetime $t$ is less than $D$=20~pc. 
Using the data in \citet{hobbs05},
we can compute the fraction of known pulsars in that sample that 
have traveled a projected distance $D$$<$20~pc at a given age $t$, 
which we call $f(t)$. If pulsars are born at a constant rate 
$R$=$2\times10^{-4}$ yr$^{-1}$ and are bright pulsar wind nebulae 
for a time $t_{\rm bright}$, then the number of nebulae in our image 
is roughly
\begin{equation}
N = R \int^{t_{\rm bright}}_{0} f(t) dt. 
\end{equation}
For $t_{\rm bright}$=100 kyr we find that only pulsars with 
$v$$>$200~km~s$^{-1}$
may have time to escape, and that out of 20 pulsars born, roughly 14 would 
remain within our image. If $t_{\rm bright}$=300 kyr, pulsars with 
$v$>70~km~s$^{-1}$ may escape, so that 29 out of 60 pulsars 
would remain. In this estimate, we have not considered that some pulsars
will be gravitationally bound to the Galactic center. If we use the 
enclosed mass in \citet{gen96}, the escape velocity from a radius of 
10 pc is only $v_{\rm esc} \approx 200$ km s$^{-1}$. Therefore, if 
$t_{\rm bright}$ is significantly longer than 100 kyr, then the fraction
of pulsars retained in the Galactic center would be larger than we estimate
(66\% of pulsars have tangential velocities smaller than 200 km s$^{-1}$,
although many of these will have larger three-dimensional velocities). 
For comparison, if we consider candidate pulsar wind nebulae to be those 
features that (1) exhibit no evidence for iron emission, and (2) are not 
obviously associated with supernova remnants already detected in the
radio, then there are also $\sim$20 of them in our image. This is consistent
with the numbers we expect if pulsars produce wind nebulae for 
100--300 kyr after their birth. 

\begin{figure*}[htp]
\centerline{\psfig{file=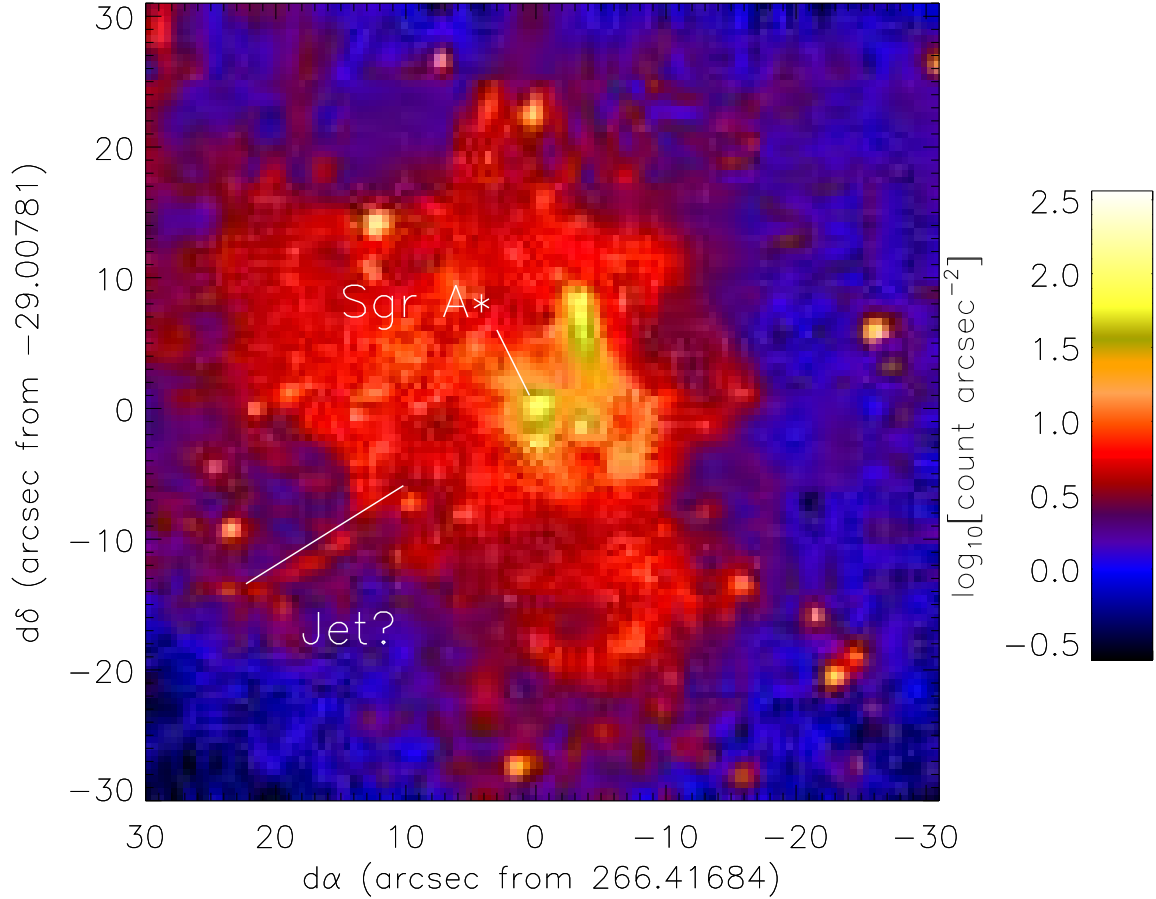,width=0.9\linewidth}}
\caption{
\chandra\ image of the central 30\arcsec\ around \sgrastar, 
at 0\farcs5 resolution. The white line pointing toward the southeast away 
from \sgrastar\ runs parallel to
a jet-like feature that was identified using our wavelet 
analysis (G359.944--0.052; source 3 in Tab.~\ref{tab:sources}). The
jet becomes noticeable at at an offset of $\approx$10\arcsec\ from 
\sgrastar.}
\label{fig:jetonly}
\end{figure*}


\subsection{A Jet from \sgrastar?}

Finally, we note that one of the X-ray features may be a jet produced
by \sgrastar, G359.944--0.052 (source 3 in Table~\ref{tab:sources}).
We highlight this source in particular in Figure~\ref{fig:jetonly}.
The object is about 15\arcsec\ long, and is unresolved along its short
axis. Our hypothesis that it is a jet is motivated by the fact that it
position angle points within 2\degree\ of \sgrastar.  From a visual
inspection of the rest of our image, we find only one candidate
feature that is similarly long and thin (located at $\alpha,\delta$=
266.37688, --29.04517), but this feature is not picked up by the
automated wavelet search based on the work of
\citet{sta00} because it is half as bright as G359.944--0.052.  
The candidate jet is found by both of our wavelet 
algorithms, so we believe this feature is unlikely to be
an artifact produced by a chance alignment of point sources, or by
Poisson variations in the diffuse background. The feature has a flat
spectrum of slope $\Gamma = -0.2\pm0.3$ and an absorption column
within 1$\sigma$ of the average toward the Galactic center, $N_{\rm H}
= (5\pm2) \times10^{22}$ cm$^{-2}$. Its luminosity is 
$L_{\rm X} = 2\times10^{32}$ \ergsec\ (2--8~keV). 

The candidate jet is not detected in the radio, with an upper limit of
$<$45 mJy at 6 cm (5 GHz).  The radio upper limit is not constraining
because the feature lies along the line-of-sight toward bright,
diffuse radio emission. The relative strength of the radio and X-ray
emission from the synchrotron-emitting jets of quasars can be
quantified using the logarithmic slope $\alpha_{\rm RX} = -
\log(S_{\rm X}/S_{\rm R}) / \log(\nu_{\rm X}/\nu_{\rm R})$, where
$S_{\rm X}$ and $S_{\rm R}$ and the X-ray and radio flux densities,
and $\nu_{\rm X}$ and $\nu_{\rm R}$ are the frequencies at which the
flux densities are computed. From the values in Table~\ref{tab:spec},
$S_{\rm X} = 0.3$ nJy at $\nu_{\rm X} = 2.4\times10^{17}$ Hz (1~keV).
Therefore, we can only constrain $\alpha_{\rm RX} < 1.1$. For
comparison, the quasars in the sample of \citet{mar05} have
$\alpha_{\rm RX}$ between 0.9 and 1.1. Therefore, one would need radio
observations with a sensitivity of $<$1 mJy to make a useful
comparison between the putative jet feature near \sgrastar\ and quasar
jets.  Further observations are warranted to determine
whether G359.944--0.052 is a jet of synchrotron-emitting particles
that is flowing away from \sgrastar.

\section{Conclusions}

There are several processes that appear to produce diffuse X-ray
features in the Galactic center: X-rays scatter off of gas in 
molecular clouds, and cause them to fluoresce; supernova shocks 
produce X-ray emitting, TeV
electrons when they propagate through low-density ISM;
and non-thermal radio filaments produce X-rays through a 
mechanism that is still not understood. 

However, we propose that the majority of the diffuse X-ray features
are pulsar wind nebulae. A significant number of pulsars are expected
at the Galactic center, but because of the high dispersion measure
along the line of sight, only two have so far been reported there in
the radio \citep{joh06}. X-ray identifications of pulsar wind nebulae
bypass the problem of the dispersion measure, and in addition are
sensitive to pulsars that are not beamed toward our line of sight.
Therefore, the identification of $\approx$20 X-ray features that are
candidate pulsar wind nebulae is consistent with a basic prediction about the
number of massive stars that have recently ended their lives in the
Galactic center. However, at the moment we lack identificaations of most of
these features at other wavelengths, so further multi-wavelength studies are 
required to confirm our hypotheses. 


\acknowledgments
We are grateful to B. Gaensler for informative conversations about
pulsar wind nebulae. 
MPM was supported by the National Aeronautics and
Space Administration through Chandra Award Number 7910613 issued by
the Chandra X-ray Observatory Center, which is operated by the
Smithsonian Astrophysical Observatory for and on behalf of the
National Aeronautics Space Administration under contract NAS8-03060.

\clearpage
\begin{landscape}
\begin{deluxetable}{clccccccccccc}
\tabletypesize{\scriptsize}
\tablecolumns{13}
\tablewidth{0pc}
\tablecaption{Diffuse Features in the Central 20 pc of the Galaxy\label{tab:sources}}
\tablehead{
\colhead{Number} & \colhead{Object} & \colhead{RA} & \colhead{DEC} & \colhead{Area} & \colhead{Offset} & \colhead{Live Time} & \colhead{Net Counts} & \colhead{$\log{P_{\rm back}}$} & \colhead{HR0} & \colhead{HR2} & \colhead{$F_{\rm X}$} & \colhead{$S_{\rm 5~GHz}$} \\
\colhead{} & \colhead{} & \multicolumn{2}{c}{(degrees, J2000)} & \colhead{arcsec$^{2}$} & \colhead{(arcmin)} & \colhead{(ks)} & \colhead{} & \colhead{} & \colhead{} & \colhead{} & \colhead{($10^{-7}$ ph cm$^{-2}$ s$^{-1}$)} & \colhead{(mJy)}
}
\startdata
 1 & G359.945-0.044 & $266.41574$ & $-29.00617$ &    22 &   0.1 & 927 & $10357.5^{+101.8}_{-101.8}$ & $<$--99.0 & $ 0.89_{-0.01}^{+0.01}$ & $ 0.11_{-0.01}^{+0.01}$ &  419.6 & $<$100 \\
 2 & G359.942-0.045 & $266.41443$ & $-29.00857$ &    28 &   0.1 & 927 & $ 3650.0^{+ 60.4}_{- 60.4}$ & --10.8 & $ 0.84_{-0.01}^{+0.01}$ & $-0.16_{-0.03}^{+0.03}$ &  132.3 & $<$30 \\
 3 & G359.944-0.052 & $266.42200$ & $-29.01091$ &    20 &   0.3 & 927 & $  695.9^{+ 62.7}_{- 62.7}$ & --11.1 & $ 1.00_{-0.13}$ & $ 0.20_{-0.05}^{+0.05}$ &   29.2 & $<$45 \\
 4 & G359.950-0.043 & $266.41788$ & $-29.00109$ &    55 &   0.4 & 927 & $ 1296.0^{+ 36.0}_{- 36.0}$ &  --8.9 & $ 0.88_{-0.02}^{+0.02}$ & $-0.01_{-0.05}^{+0.05}$ &   50.6 & 210$\pm$30 \\
 5 & G359.933-0.039 & $266.40372$ & $-29.01335$ &    15 &   0.8 & 927 & $  200.4^{+ 32.6}_{- 31.6}$ &  --3.4 & $ 0.29_{-0.77}^{+0.28}$ & $-0.03_{-0.12}^{+0.10}$ &    8.1 & $<$3 \\
 6 & G359.956-0.052 & $266.43002$ & $-29.00123$ &    10 &   0.8 & 927 & $  295.5^{+ 41.0}_{- 40.2}$ &  --3.1 & $ 1.00_{-0.55}$ & $ 0.14_{-0.09}^{+0.08}$ &   12.7 & $<$9 \\
 7 & G359.933-0.037 & $266.40140$ & $-29.01229$ &    13 &   0.9 & 927 & $  542.9^{+ 44.0}_{- 44.0}$ & --17.2 & $ 0.72_{-0.08}^{+0.06}$ & $ 0.09_{-0.05}^{+0.05}$ &   22.8 & $<$5 \\
 8 & G359.941-0.029 & $266.39868$ & $-29.00148$ &    17 &   1.0 & 927 & $  430.5^{+ 41.7}_{- 40.9}$ & --11.7 & $ 1.00_{-0.32}$ & $ 0.19_{-0.05}^{+0.05}$ &   18.7 & $<$2 \\
 9 & G359.925-0.051 & $266.41000$ & $-29.02688$ &    19 &   1.2 & 927 & $  427.7^{+ 41.0}_{- 40.6}$ &  --7.5 & $ 1.00_{-0.42}$ & $ 0.06_{-0.06}^{+0.06}$ &   18.1 & $<$2 \\
10 & G359.964-0.053 & $266.43475$ & $-28.99465$ &    76 &   1.2 & 927 & $ 3424.4^{+ 58.5}_{- 58.5}$ & --41.9 & $ 0.82_{-0.01}^{+0.01}$ & $ 0.01_{-0.02}^{+0.02}$ &  151.1 & 11$\pm$3 \\
11 & G359.965-0.056 & $266.43893$ & $-28.99563$ &    29 &   1.4 & 927 & $  568.7^{+ 60.7}_{- 60.4}$ &  --6.8 & $ 1.00_{-0.22}$ & $-0.12_{-0.09}^{+0.08}$ &   22.4 & $<$2 \\
12 & G359.921-0.052 & $266.40854$ & $-29.03010$ &    12 &   1.4 & 927 & $  169.2^{+ 28.4}_{- 26.9}$ &  --3.6 & $ 1.00_{-0.40}$ & $ 0.29_{-0.09}^{+0.08}$ &    7.5 & $<$1 \\
13 & G359.962-0.062 & $266.44254$ & $-29.00101$ &    26 &   1.4 & 927 & $  523.9^{+ 59.0}_{- 59.0}$ &  --1.1 & $ 1.00_{-0.15}$ & $-0.07_{-0.10}^{+0.09}$ &   20.3 & 2$\pm$1 \\
14 & G359.959-0.027 & $266.40738$ & $-28.98546$ &    34 &   1.4 & 927 & $  773.7^{+ 27.8}_{- 27.8}$ & --32.7 & $ 1.00_{-0.22}$ & $ 0.08_{-0.04}^{+0.04}$ &   33.6 & 20$\pm$3 \\
15 & G359.971-0.038 & $266.42542$ & $-28.98094$ &   148 &   1.7 & 927 & $ 1786.7^{+ 42.3}_{- 42.3}$ & --19.8 & $ 0.64_{-0.07}^{+0.06}$ & $ 0.01_{-0.04}^{+0.04}$ &   71.4 & $<$10  \\
16 & G359.969-0.033 & $266.41895$ & $-28.97952$ &    17 &   1.7 & 927 & $  207.4^{+ 31.5}_{- 30.6}$ &  --5.6 & $ 1.00_{-0.47}$ & $ 0.28_{-0.08}^{+0.07}$ &    9.0 & $<$1 \\
17 & G359.921-0.030 & $266.38730$ & $-29.01968$ &    30 &   1.7 & 927 & $  391.8^{+ 39.1}_{- 38.9}$ & --12.7 & $ 1.00_{-0.41}$ & $-0.03_{-0.07}^{+0.06}$ &   15.7 & $<$1 \\
18 & G359.915-0.061 & $266.41370$ & $-29.04057$ &    22 &   2.0 & 927 & $  192.3^{+ 29.5}_{- 28.3}$ & --15.2 & $-9.00$ & $ 0.54_{-0.05}^{+0.04}$ &   10.0 & 1.8$\pm$0.7 \\
19 & G359.983-0.040 & $266.43433$ & $-28.97174$ &    35 &   2.4 & 927 & $  139.6^{+ 36.5}_{- 36.0}$ &  --0.5 & $ 0.51_{-0.19}^{+0.12}$ & $-0.11_{-0.33}^{+0.23}$ &    5.4 & $<$1 \\
20 & G359.904-0.047 & $266.39368$ & $-29.04198$ &    32 &   2.4 & 927 & $  296.5^{+ 35.3}_{- 34.6}$ &  --6.7 & $ 1.00_{-0.39}$ & $ 0.28_{-0.06}^{+0.05}$ &   13.2 & $<$0.2 \\
21 & G359.977-0.076 & $266.46542$ & $-28.99490$ &    26 &   2.7 & 927 & $  164.4^{+ 29.9}_{- 29.0}$ &  --1.6 & $-0.35_{-0.39}^{+0.28}$ & $ 0.19_{-0.13}^{+0.11}$ &    6.5 & $<$4 \\
22 & G359.970-0.008 & $266.39569$ & $-28.96591$ &    30 &   2.7 & 927 & $  640.7^{+ 47.1}_{- 46.7}$ & --32.6 & $ 1.00_{-0.26}$ & $ 0.11_{-0.04}^{+0.04}$ &   28.1 & 1.2$\pm$0.4 \\
23 & G359.899-0.065 & $266.40851$ & $-29.05584$ &    30 &   2.9 & 927 & $  222.2^{+ 30.1}_{- 28.6}$ & --10.7 & $ 0.41_{-0.32}^{+0.18}$ & $ 0.42_{-0.06}^{+0.05}$ &   11.0 & 6$\pm$1 \\
24 & G359.897-0.023 & $266.36603$ & $-29.03623$ &    42 &   3.2 & 927 & $  233.1^{+ 32.9}_{- 31.3}$ &  --2.3 & $ 0.45_{-0.15}^{+0.11}$ & $-0.03_{-0.11}^{+0.10}$ &    9.2 & $<$0.4 \\
25 & G359.889-0.081 & $266.41791$ & $-29.07335$ &   432 &   3.9 & 927 & $ 5364.7^{+ 73.2}_{- 73.2}$ & $<$-99.0 & $ 0.45_{-0.12}^{+0.09}$ & $ 0.45_{-0.01}^{+0.01}$ &  281.9 & \nodata \\
26 & G0.014-0.054 & $266.46631$ & $-28.95223$ &   260 &   4.2 & 927 & $ 1551.1^{+ 39.4}_{- 39.4}$ & --28.4 & $-0.04_{-0.89}^{+0.39}$ & $ 0.54_{-0.02}^{+0.02}$ &   78.1 & $<$0.6 \\
27 & G0.008-0.015 & $266.42432$ & $-28.93708$ &    51 &   4.3 & 927 & $  223.2^{+ 34.2}_{- 32.9}$ &  --3.1 & $ 0.24_{-0.34}^{+0.21}$ & $-0.11_{-0.13}^{+0.11}$ &    9.2 & 1.6$\pm$0.5 \\
28 & G0.021-0.051 & $266.46741$ & $-28.94492$ &   190 &   4.6 & 927 & $ 1326.7^{+ 36.4}_{- 36.4}$ & --27.8 & $ 1.00_{-0.69}$ & $ 0.61_{-0.02}^{+0.02}$ &   69.2 & \nodata \\
29 & G0.032-0.056 & $266.47849$ & $-28.93840$ &   429 &   5.3 & 927 & $ 2305.2^{+ 48.0}_{- 48.0}$ & --30.8 & $ 1.00_{-0.21}$ & $ 0.02_{-0.04}^{+0.04}$ &  100.9 & \nodata  \\
30 & G0.029-0.080 & $266.50043$ & $-28.95345$ &   838 &   5.5 & 927 & $ 1398.2^{+ 37.4}_{- 37.4}$ & --18.3 & $ 1.00_{-0.74}$ & $ 0.49_{-0.04}^{+0.04}$ &   73.0  & \nodata \\
31 & G0.039-0.077 & $266.50354$ & $-28.94300$ &   333 &   6.0 & 927 & $ 2424.8^{+ 49.2}_{- 49.2}$ & --33.5 & $ 0.59_{-0.10}^{+0.07}$ & $ 0.29_{-0.02}^{+0.02}$ &  118.4  & \nodata \\
32 & G0.062+0.010 & $266.43198$ & $-28.87782$ &  1187 &   7.8 & 927 & $ 1566.5^{+ 39.6}_{- 39.6}$ & --14.7 & $ 0.72_{-0.06}^{+0.05}$ & $ 0.06_{-0.06}^{+0.05}$ &   75.9  & \nodata \\
33 & G0.097-0.131 & $266.59061$ & $-28.92188$ &  4181 &  10.5 & 563 & $ 5246.9^{+ 72.4}_{- 72.4}$ & $<$-99.0 & $ 0.12_{-0.44}^{+0.25}$ & $ 0.21_{-0.02}^{+0.02}$ &  529.0  & \nodata \\
34 & G0.116-0.111 & $266.58212$ & $-28.89475$ &  2257 &  11.0 & 574 & $ 1291.3^{+ 35.9}_{- 35.9}$ &  --7.3 & $ 0.18_{-0.41}^{+0.23}$ & $ 0.24_{-0.07}^{+0.06}$ &  121.6  & \nodata
\enddata
\tablecomments{The object name is constructed from the Galactic coordinates
in degrees. The position is given in degrees as the right ascension and 
declination, at the epoch J2000. The area is that of the extraction
region in Figures 1--3. The offset is from \sgrastar. The live time
is the amount of time for which the target was in the field of view of the
detector. The net counts are in the 0.5--8.0~keV band, and the 
uncertainties are 1$\sigma$. $P_{\rm back}$ is 
the probability that the spectrum of the 
feature is the same as that of the background, estimated via a KS test. The
hardness ratios are defined as $(h-s)/(h+s)$. For HR0, $h$ is the 2.0--3.3~keV
band, and $s$ is the 0.5--2.0~keV band. For HR2, $h$ is the 4.7--8.0~keV
band, and $s$ is the 3.3--4.7~keV band. HR1 is not used. The flux is
the observed value from the 0.5--8.0~keV band. The radio fluxes are taken 
from the 6 cm (5 GHz) image in Figure~\ref{fig:radiob}.}
\end{deluxetable}
\clearpage
\end{landscape}


\begin{thebibliography}{0}
\bibitem[Aharonian \etal(2004)]{aha04} Aharonian, F. \etal\ 2004, \aap, 
  425, L13
\bibitem[Arabadjis \etal(2004)]{aba04} Arabadjis, J. S., Bautz, M. W., \&
  Arabadjis, G. 2004, \apj, 617, 303
\bibitem[Arnaud \etal(1996)]{arn96} Arnaud, K.A., 1996, Astronomical Data 
 Analysis Software and Systems V,  eds. Jacoby G. and Barnes J., p17, ASP 
 Conf. Series volume 101
\bibitem[Arzoumanian, Chernoff, \& Cordes(2002)]{arz02} 
  Arzoumanian Z., Chernoff D. F., \& Cordes J. M., 2002, ApJ, 568, 289
\bibitem[Baganoff \etal(2003)]{bag03} Baganoff, F. K. \etal\ 2003, 
  \apj, 591, 891
\bibitem[Becker \& Tr\"{u}mper(1997)]{bt97} Becker, W. \& Tr\"{u}mper, J.
  1997, \aap, 326, 682
\bibitem[Bertin \& Arnouts(1996)]{ba96}
  Bertin, E. \& Arnouts, S. 1996, \aap, 117, 393
\bibitem[Brazier \& Johnston(1999)]{bj99} Brarzier, K. T. S. \& Johnston, S.
  1999, \mnras, 305, 671
\bibitem[Bykov(2002)]{byk02} Bykov, A. M. 2002, \aap, 390, 327
\bibitem[Cheng \etal(2004)]{ctw04} Cheng, K. S., Taam, R. E., \& Wang, W.
  2004, \apj, 617, 480
\bibitem[Cheng \etal(2006)]{ctw06} Cheng, K. S., Taam, R. E., \& Wang, W.
  2006, \apj, 641, 427 
\bibitem[Cordes \& Lazio(1997)]{cl97} Cordes, J. M. \& Lazio, T. J. W. 
  1997, \apj, 475, 557
\bibitem[Faucher-Gigu\`{e}re \& Kaspi(2006)]{fgk06} 
  Faucher-Gigu\`{e}re, C.-A. \& Kaspi, V. M. 2006, \apj, 643, 332
\bibitem[Figer \etal(2004)]{fig04} Figer, D. F., Rich, R. M., Kim, S. S., 
  Morris, M., \& Serabyn, E. 2004, \apj, 610, 317
\bibitem[Frail \& Scharringhausen(1997)]{fs97} Frail, D. A., \& 
  Scharringhausen, B. R. 1997, \apj, 480, 364
\bibitem[Freeman \etal(2002)]{free02} Freeman, P. E., Kashyap, V., 
  Rosner, R., \& Lamb, D. Q. 2002, \apjs, 138, 185
\bibitem[Gaensler \& Slane(2006)]{gs06} Gaensler, B. M. \& Slane, P. O.
  2006, \araa, 44, 17
\bibitem[Gaensler \etal(2000)]{gae00} Gaensler, B. M., Stappers, B. W., 
  Frail, D. A., Moffat, D. A., Johnston, S., \& Chatterjee, S. 2000, 
  \mnras, 318, 58\
\bibitem[Gaensler \etal(2003)]{gae03} Gaensler, B. M., Schulz, N. S., 
  Kaspi, V. M., Pivovaroff, M. J., \& Becker, W. E. 2003, \apj, 588, 441
\bibitem[Genzel \etal(1996)]{gen96} Genzel, R., Thatte, N., Krabbe, A., 
  Kroker, H., \& Tacconi-Garman, L. E. 1996, \apj, 472, 153
\bibitem[Gonzales \etal(2006)]{gon06} Gonzales, M. E., Kaspi, V. M., 
  Pivovaroff, M. J. \& Gaensler, B. M. 2006, \apj, 652, 569
\bibitem[Gotthelf, Halpern, \& Dodson(2002)]{ghd02} Gotthelf, E. V., 
  Halpern, J. P., \& Dodson, R. 2002, \apj, 567, L125
\bibitem[Gotthelf \& Kaspi(1998)]{gk98} Gotthelf, E. V. \& Kaspi, V. .M.
  1998, \apj, 497, 429
\bibitem[G\"{u}sten \etal(1981)]{gwp81} G\"{u}sten, R., Walmsley, C. M., \&
  Pauls, T. 1981, \aap, 103, 197
\bibitem[Heger \etal(2003)]{heg03} Heger, A., Fryer, C. L., Woosley, S. E., 
  Langer, N., \& Hartmann, D. H. 2003, \apj, 591, 288
\bibitem[Ho \etal(1985)]{ho85} Ho, P. T. P., Jackson, J. M., Barret, A. H., \&
  Armstrong, T. J. 1985, \apj, 288, 575
\bibitem[Hobbs \etal(2005)]{hobbs05} Hobbs, G., Lorimer, D. R., 
  Lyne, A. G., \& Kramer, M. 2005, \mnras, 360, 974 
\bibitem[Johnston \etal(2006)]{joh06} Johnston, S., Kramer, M., Lorimer, D. R.,
  Lyne, A. G., McLaughlin, M., Klein, B., \& Manchester, R. N. 2006, \mnras,
  373, L6
\bibitem[Krabbe et al.(1995)]{kra95} Krabbe, A. \etal\ 1995, \apj, 447, L95
\bibitem[Kroupa(2002)]{krou02} Kroupa, P. 2002, Science, 295, 82
\bibitem[Koyama \etal(1996)]{koy96} Koyama, K., Maeda, Y., Sonobe, T., 
  Takeshima, T., Tanaka, Y., \& Yamauchi, S. 1996, \pasj, 48, 249
\bibitem[LaRosa \etal(2005)]{lar05} LaRosa, R. N., Brogan, C. L., 
  Shore, S. M., Lazio, T. J. W., Kassim, N. E., \& Nord, M. E. 2005, \apj,
  626, L23
\bibitem[Li \etal(2004)]{li04} Li, J., Kastner, J. H., Prigozhin, G. Y., 
  Schulz, N. S., Feigelson, E. D., \& Getman, K. V. 2004, \apj, 610, 1204
\bibitem[Longair(1994)]{lon94} Longair, M. S. 1994, {\it High Energy 
  Astrophysics}, volume 2, 2nd ed., Cambridge University Press
\bibitem[Lu \etal(2003)]{lwl03} Lu, F. J., Wang, Q. D., \& Lang, C. C. 2003, 
  \aj, 126, 319
\bibitem[Lyne \etal(1998)]{lyn98} Lyne, A. G. \etal\ 1998, \mnras, 295, 743
\bibitem[Maeda \etal(2002)]{mae02} Maeda, Y. \etal\ 2002, \apj, 570, 671
\bibitem[Manchester \etal(2005)]{man05} Manchester, R. N., Hobbs, G. B., 
  Teoh, A., \& Hobbs, M. 2005, \aj, 129, 1993
\bibitem[Marshall \etal(2005)]{mar05} Marshall, H. M. \etal\ 2005, 
  \apjs, 156, 13
\bibitem[McGowan \etal(2006)]{mcg06} McGowan, K. E., Vestrand, W. T., 
  Kennea, J. A., Zane, S., Cropper, M., \& C\'{o}rdova, F. A., 2006, 
  \apj, 647, 1300
\bibitem[Morris \etal(2003)]{mor03} Morris, M. \etal\ 2003, ANS, 324, 167a
\bibitem[Morris \& Serabyn(1996)]{ms96} Morris, M. \& Serabyn, E. 1996, 
 \araa, 34, 645
\bibitem[Muno \etal(2003)]{m-cat} Muno, M. P. \etal\ 2003, \apj, 589, 225
\bibitem[Muno \etal(2004a)]{m-diff} Muno, M. P. \etal\ 2004a, \apj, 613, 326
\bibitem[Muno \etal(2004b)]{m-ps} Muno, M. P. \etal\ 2004b, \apj, 613, 1179
\bibitem[Muno \etal(2007)]{mun07} Muno, M. P., Baganoff, F. K., Brandt, W. N., 
  Park, S. \& Morris, M. R. 2007, \apj, 656, L69
\bibitem[Murakami \etal(2001a)]{mkm01} Murakami, H., Koyama, K., \& Maeda, 
  Y. 2001, \apj, 558, 687 
\bibitem[Murakami \etal(2001b)]{mur01b} Murakami, H., Koyama, K., 
  Tsujimoto, M., Maeda, Y., \& Sakano, M. 2001b, \apj, 550, 297
\bibitem[Nord \etal(2004)]{nor04} Nord, M. E., Lazio, T. J. W., Kassim, N. E.,
  Hyman, S. D., Larosa, T. N., Brogan, C. L., \& Duric, N. 2004, \aj, 128, 1646
\bibitem[Oka \etal(1998)]{oka98} Oka, T., Hasegawa, T., Hayashi, M., Handa, T.,
  \& Sakamoto, S. 1998, \apj, 493, 730
\bibitem[Park \etal(2004)]{par04} Park, S., Muno, M. P., Baganoff, F. K., 
  Maeda, Y., Morris, M., Howard, C., Bautz, M. W., \& Garmire, G. P. 2004, 
  \apj, 603, 548
\bibitem[Park \etal(2005)]{par05} Park, S. \etal\ 2005, \apj, 631, 964
\bibitem[Paumard \etal(2006)]{pau06} Paumard, T. \etal\ 2006, \apj, 
  643, 1011
\bibitem[Pfahl \& Loeb(2004)]{pl04} Pfahl, E. \& Loeb, A. 2004, \apj, 
  615, 253
\bibitem[Revnivtsev \etal(2004)]{rev04} Revnivtsev, M. G. \etal\ 2004, 
  \aap, 425, L49
\bibitem[Revnivtsev \etal(2007)]{rvs07} Revnivtsev, M., 
  Vikhlinin, A., \& Sazonov, S. 2007, submitted to \aap, astro-ph/0611952
\bibitem[Romani \etal(2005)]{rom05} Romani, R. W., Ng, C. Y., Dodson, R., \&
  Brisker, W. 2005, \apj, 631, 480
\bibitem[Rybicki \& Lightman(1979)]{rl79} Rybicki, G., \& Lightman, A.
  {\it Radiative Processes in Astrophysics}, 1979, 
  John Wiley \& Sons, Inc.
\bibitem[Sakano \etal(2003)]{sak03} Sakano, M., Warwick, R. S., 
  Decourchelle, A., \& Predehl, P. 2003, \mnras, 340, 747
\bibitem[Sakano \etal(2004)]{sak04} Sakano, M., Warwick, R. S., 
  Decourchelle, A., \& Predehl, P. 2004, \mnras, 350, 129
\bibitem[Scoville \etal(2003)]{sco03} Scoville, N. Z., Stolovy, S. R., 
  Reike, M., Christopher, M., \& Yusef-Zadeh, F. 2003, \apj, 594, 294
\bibitem[Skrutskie \etal(2006)]{skr06} Skrutskie, M. F. \etal\ 2006, \aj,
  131, 1163
\bibitem[Starck \etal(2000)]{sta00}
  Starck, J.-L., Bijauoi, A., Valtchanov, I., \& Murtagh, F. 2000, \aap, 
  147, 139
\bibitem[Sunyaev \etal(1993)]{smp93} Sunyaev, R. A., Markevitch, M., \& 
  Pavlinsky, M. 1993, \apj, 407, 606 
\bibitem[Valinia \etal(2000)]{val00} Valinia, A., Tatischeff, V., Arnaud, K., 
  Ebisawa, K., \& Ramaty, R. 2000, \apj, 543, 733
\bibitem[Wang \etal(2006)]{wlg06} Wang, Q. D., Lu, F. J., \& Gotthelf, 
  E. V. 2006, \mnras, 367, 937 
\bibitem[Wang \etal(2002b)]{wll02b} Wang, Q. D., Lu, F. J., \& Lang, C. C.
  2002b, \apj, 581, 1148
\bibitem[Wang \etal(2005)]{wjc05} Wang, W., Jiang, Z. J., \& Cheng, K. S.
  2005, \mnras, 358, 263
\bibitem[Weisskopf \etal(2002)]{wei02} Weisskopf, M. C., Brinkman, B., 
  Canizares, C., Garmire, G., Murray, S., \& van Speybroeck, L. P. 2002, 
  \pasp, 114, 1
\bibitem[Vink \etal(2006)]{vink06} Vink, J., Bleeker, J., van der Heyden, K., 
  Bykov, A., Bamba, A. \& Yamazaki, R. 2006, submitted to \apj, 
  astro-ph/0607307
\bibitem[Yusef-Zadeh \& Morris(1987)]{yzm87} Yusef-Zadeh, F. \& Morris, M. 
  1987, \apj, 320, 545
\bibitem[Yusef-Zadeh \etal(2005)]{yz05} Yusef-Zadeh, F., Wardle, M., Muno, M.,
  Law, C., \& Pound, M. 2005, AdSpR, 35, 1074
\bibitem[Yusef-Zadeh \etal(2004)]{yzhc04} Yusef-Zadeh, F., Hewitt, J. W., 
  \& Cotton, W. 2004, \apjs, 155, 421
\end{thebibliography}
\end{document}